\def\rO{{\rm O}}
\def\ro{{\rm o}}

\def\Wvel{{\bf W}^{\rm flow}}
\def\Wtem{{\bf W}^{\rm temp}}
\def\Uvel{{\bf U}^{\rm flow}}
\def\Utem{{\bf U}^{\rm temp}}
\def\tr{{\rm tr\,}}
\documentclass[12pt]{article}
\usepackage{epsfig,amsfonts,amssymb}
\begin{document}
\begin{center}
{\bf\large ON STABILITY OF ROLLS\\
NEAR THE ONSET OF CONVECTION\\
IN A LAYER WITH STRESS-FREE BOUNDARIES}

\bigskip
\renewcommand{\thefootnote}{\fnsymbol{footnote}}
{\large Olga Podvigina}\footnote{Email: olgap@mitp.ru}

\bigskip
\noindent
International Institute of Earthquake Prediction Theory\\
and Mathematical Geophysics,\\
84/32 Profsoyuznaya St, 117997 Moscow, Russian Federation;\\

\bigskip
UNS, CNRS, Laboratoire Cassiop\'ee, Observatoire de la C\^ote d'Azur\\
BP 4229, 06304 Nice Cedex 4, France\\

\bigskip
\end{center}

\begin{abstract}
We consider a classical problem of linear stability of convective rolls
in a plane layer with stress-free horizontal boundaries near the onset of
convection. The problem has been studied by a number of authors,
who have shown that rolls of wave number $k$ are unstable with respect to
perturbations of different types, if some inequalities relating $k$ and
the Rayleigh number $R$ are satisfied. The perturbations involve a large-scale
mode. Certain asymptotic dependencies between wave numbers of the mode and
overcriticality are always assumed in the available proofs of instability.
We analyse the stability analytically following the approach of Podvigina (2008)
without making a priori assumptions concerning asymptotic relations
between small parameters characterising the problem.
Instability of rolls to short-scale modes is also considered.
Therefore, our analytical results on stability to space-periodic perturbations
are exhaustive; they allow to identify the areas
in the $(k,R)$ plane, where convective rolls are stable near the onset.
The analytical results are compared with numerical
solutions to the eigenvalue problem determining stability of rolls.
\end{abstract}


\pagebreak
\section{Introduction}

We consider Boussinesq convection in a plane horizontal layer heated from below
with stress-free horizontal boundaries. For small Rayleigh
numbers $R$, i.e. for small temperature differences between the lower and upper
boundaries, the fluid is not moving and the heat is transported by thermal
diffusion only. As $R$ exceeds the critical value $R^s$, the fluid motion sets
in. The motion takes the form of rolls. We denote by $k_s$ the horizontal
wave number of the mode becoming unstable the first. (By a horizontal wave
number we understand the length of the horizontal component of the wave vector.)
In this paper we study analytically and numerically stability of rolls
of a horizontal wave number close to $k_s$ for the Rayleigh number slightly
above $R^s$.

Instability of rolls in a convective layer was studied analytically by a number of authors.
An instability of rolls specific for stress-free boundaries is known,
which does not occur in a layer whose one or both horizontal boundaries are
rigid. Its presence relies on existence of a slowly decaying large-scale mode.
Zippelius and Siggia (1982, 1983) were the first to study the instability
of this kind. In the leading order the unstable mode is a sum of a large-scale
mode and of two short-scale modes with wave vectors close to the one of the
perturbed rolls.
In the study of stability of rolls in a rotating convective layer it was
called {\it small angle} instability (Cox and Matthews 2000).

By deriving amplitude equations, Zippelius and Siggia obtained sufficient
conditions for instability of rolls, in particular, they found that for
$P<0.782$ ($P$ denoting the Prandtl number) no stable rolls existed near the
onset. Their results were questioned by Busse and Bolton (1984), who found
boundaries for instabilities of rolls by direct calculations of the unstable
mode, and claimed that no stable rolls were present near the onset only for
$P<0.543$. Their results were confirmed by Bernoff (1994), who studied
instability of rolls applying Ginzburg-Landau equations.

The conflict between the results of Zippelius and Siggia (1982, 1983) and
Busse and Bolton (1984) was resolved by Mielke (1997), who studied stability of
rolls by means of Lyapunov-Schmidt reduction and showed that instability
boundaries had been found in these papers for different unstable modes.
The problem of stability of rolls involves five small parameters: two
wave numbers of the large scale mode, overcriticality, the difference between
the wave number of perturbed rolls and the critical wave number, and the growth
rate (depending on the first four). Zippelius and Siggia postulated
asymptotic relations between the parameters, different from those postulated by
Busse and Bolton, and hence different instability modes were examined.

However, in all these studies some asymptotic relations between the small
parameters of the problem were assumed, and thus stability only to selected
types of perturbations was studied. Hence, only {\it instability} of rolls
was proven (as it is discussed in introduction and conclusion in
Mielke (1997) and also section 3 of Bernoff (1994). The question, whether other
unstable modes corresponding to other asymptotic scalings exist, remains open.
Another question asked by Busse and Bolton (1984) and Bernoff (1994)
and not answered by previous studies is whether enough terms
of asymptotic expansions were taken into account.
For four independent small parameters this is a hard question!

In the present paper both difficulties are resolved. We do not assume any asymptotic
relations between the small parameters, hence instability with respect to all
perturbations of the small-angle type is examined. An unstable mode is
represented as a series in small parameters, with estimates for the remainder.
(Estimates for omitted terms were not given before.)
A condition for instability has the form of inequalities. We demonstrate
that in these inequalities the omitted terms are asymptotically smaller
than the terms retained in the analysis.

We also study instability to perturbations of a different kind,
which are in the leading order convective rolls with the horizontal wave number
close to the critical one, and a finite angle between wave vectors of the
perturbed rolls and of the perturbation, which we call a {\it finite angle}
instability.

Furthermore, we show that stability to all considered perturbations implies
stability to a much wider class of doubly periodic in horizontal directions
perturbations. We show, that the domain of the linearisation operator, acting
on vector fields satisfying the assumed boundary conditions on the horizontal
boundaries and doubly periodic in horizontal directions, splits into a direct
sum of invariant subspaces. Thus investigation of instability is reduced to
detection of instability modes in these invariant subspaces.
Any instability mode in such a subspace is either of a small angle or
finite angle type. Therefore, our study
of instabilities of rolls with respect to perturbations which are doubly
periodic in some directions in the $(x,y)$ plane is complete:
we demonstrate {\it stability} of rolls for which instability is not
detected in our analysis. This is also a novel feature of our investigation:
all previous papers focused exclusively on instability.

We estimate analytically the asymptotics of the most unstable mode and
its growth rate on different parts of the $(k,R)$ plane and calculate
the instability boundaries estimating the orders of the neglected terms
(which was not yet done in literature).

We are using notation and some results of Podvigina (2008), referred to as
OP2008, where instability of flows in a rotating convective layer with
stress-free boundaries was studied.

Stability of rolls in a layer with stress-free boundaries was studied
by direct computations of dominant eigenvalues by Bolton and Busse (1985).
Their numerical results agree well with the theoretical predictions of Busse
and Bolton (1984), in particular, they found that for $P=0.71$ rolls are stable
near the onset. However, Mielke (1997) claimed that stable rolls near the onset
for $P<0.782$ are absent; he did not comment on the disagreement with
the numerical results of Bolton and Busse (1985).
In contrast to (Bolton and Busse 1985), our computations of stability modes
indicate that rolls are unstable at the onset for $P<0.782$, albeit
in a small neighbourhood of the point $(k_s,R^s)$ on the $(k,R)$ plane.
For $P$ decreasing below $0.782$, the right boundary of the area of stable
rolls slowly moves to the left, away from the point $(k_s,R^s)$.

\section{The onset of convection}

Boussinesq thermal convection satisfies the Navier-Stokes equation
\begin{equation}
{\partial{\bf v}\over\partial t}={\bf v}\times(\nabla\times{\bf v})
+P\nabla^2{\bf v}+PR\theta{\bf e}_z-\nabla p,
\label{nst}\end{equation}
the incompressibility condition
\begin{equation}
\nabla\cdot{\bf v}=0
\label{inc}\end{equation}
and the heat transfer equation
\begin{equation}
{\partial\theta\over\partial t}=-({\bf v}\cdot\nabla)\theta+v_z+\nabla^2\theta
\label{heat}\end{equation}
where $\bf v$ is the flow velocity and $\theta$ is the deviation
of the flow temperature from the linear profile.
$R$ and $P$ are dimensionless parameters, the Rayleigh and Prandtl numbers,
respectively. Stress-free horizontal boundaries held
at fixed temperature are assumed:
\begin{equation}
{\partial v_x\over\partial z}={\partial v_y\over\partial z}=v_z=0,
\qquad\theta=0\qquad\hbox{at }z=0,1.
\label{bouc}\end{equation}

The trivial solution $({\bf v},\theta)=({\bf 0},0)$
describing pure thermal conduction loses
stability to perturbations of wave number $k$ at $R=R_c(k)$, where
\begin{equation}
R_c(k)=a^3k^{-2},\quad a=k^2+\pi^2.
\label{rcs}\end{equation}
The critical horizontal wave number $k_s$ for the onset of convection is
$\pi/\sqrt{2}$, the respective critical value $R^s=27\pi^4/4$.

We employ four-dimensional vectors
\begin{equation}
{\bf W}\equiv(\Wvel,\Wtem)=({\bf v},\theta).
\label{fcom}\end{equation}

For a Rayleigh number slightly above the critical value,
\begin{equation}
R=R_c+\varepsilon^2,
\label{r2c}\end{equation}
a solution to (\ref{nst})-(\ref{bouc}) can be represented as a power series
\begin{equation}
{\bf U}=\sum_{j=1}^\infty\varepsilon^j{\bf U}_j.
\label{seri}\end{equation}
The first two terms of the solution representing rolls are
\begin{equation}
\label{v1}
{\bf U}_1=b \left(
\begin{array}{c}
-\pi k^{-1}\cos\pi z\sin kx\\
0\\
\sin\pi z\cos kx\\
a^{-1}\sin\pi z\cos kx
\end{array}
\right)
\end{equation}
(${\bf U}_1$ is an eigenvector of the linearisation of (\ref{nst})-(\ref{heat})~),
\begin{equation}
\label{v2}
{\bf U}_2=b^2 \left(
\begin{array}{c}
0\\
0\\
0\\
-(8\pi a)^{-1}\sin 2\pi z\\
\end{array}
\right),
\end{equation}
where
\begin{equation}
8a=R_cb^2.
\label{ampA}\end{equation}

\section{Invariant subspace}

We study stability of rolls of wave number $k$, which is close to the critical
one;
\begin{equation}
\label{alph}
\alpha=k-k_s
\end{equation}
is thus a small parameter.

Stability of (\ref{seri}) is controlled by eigenvalues of the linear
operator $L$, a linearisation of (\ref{nst})-(\ref{heat}) near
the steady state. The operator can be expanded in a series
\begin{equation}
L=\sum_{j=0}^\infty\varepsilon^j L_j.
\label{seriL}\end{equation}
Here the first term is
\begin{equation}
L_0({\bf v},\theta)=(P\nabla^2{\bf v}+PR_c\theta{\bf e}_z-\nabla p,
v_z+\nabla^2\theta).
\label{L0}\end{equation}

We consider the eigenvalue problem
\begin{equation}
L{\bf W}=\lambda{\bf W}.
\label{eig}\end{equation}
In OP2008 a three-dimensional invariant
subspace of $L$ was considered, and the problem of stability of
rolls was reduced to the analysis of eigenvalues of $L$ in this subspace.
Denote by ${\bf W}_j$, $j=1,2,3$, a basis in this subspace and by
$\cal A$ the matrix $A_{ij}$ of the restriction of $L$ on the subspace:
\begin{equation}
L{\bf W}_j=\sum_{i=1}^3A_{ij}{\bf W}_i,\quad j=1,2,3.
\label{ivs}\end{equation}
We expand the basis and the matrix in power series in $\varepsilon$:
\begin{equation}
{\bf W}_j=\sum_{l=0}^\infty\varepsilon^l{\bf W}_{j,l},
\label{seriW}\end{equation}
\begin{equation}
A_{ij}=\sum_{l=0}^\infty\varepsilon^l A_{ij,l}.
\label{seriA}\end{equation}
The coefficients in the series (\ref{seriW}) and (\ref{seriA})
depend on $\delta_x$ and $\delta_y$, which are small perturbations
of the horizontal component of the wave vector $(k,0,\pi)$:
$\delta_x\ll k$ and $\delta_y\ll k$. (The two small parameters are involved
in the definition of the three-dimensional invariant subspace of $L$, see OP2008.)

Vector fields ${\bf W}_{j,0}$ are eigenfunctions of $L_0$:
\begin{equation}
\label{bubuW}
L_0{\bf W}_{j,0}=\lambda_{j,0}{\bf W}_{j,0},
\end{equation}
the leading order approximations of ${\bf W}_{j,0}$ in $\delta_x$ and $\delta_y$
were derived in OP2008:
\begin{equation}
\label{W01}
{\bf W}_{1,0}= \left(
\begin{array}{c}
-\delta_y\sin(\delta_x x+\delta_y y)\\
\delta_x\sin(\delta_x x+\delta_y y)\\
0\\
0
\end{array}
\right),
\end{equation}
\begin{equation}
\label{W02}
{\bf W}_{2,0}= \left(
\begin{array}{c}
-\pi s_+k_+^{-1}\cos\pi z\sin((k+\delta_x)x+\delta_y y)\\
-\pi\delta_y k_+^{-2}\cos\pi z\sin((k+\delta_x)x+\delta_y y)\\
\sin\pi z\cos((k+\delta_x)x+\delta_y y)\\
a_+^{-1}\sin\pi z\cos((k+\delta_x)x+\delta_y y)
\end{array}
\right)+\rO((k_+^2-k^2)^2,\alpha(k_+^2-k^2)),
\end{equation}
\begin{equation}
\label{W03}
{\bf W}_{3,0}=\left(
\begin{array}{c}
-\pi s_-k_-^{-1}\cos\pi z\sin((k-\delta_x)x-\delta_y y)\\
\pi\delta_y k_-^{-2}\cos\pi z\sin((k-\delta_x)x-\delta_y y)\\
\sin\pi z\cos((k-\delta_x)x-\delta_y y)\\
a_-^{-1}\sin\pi z\cos((k-\delta_x)x-\delta_y y)
\end{array}
\right)+\rO((k_-^2-k^2)^2,\alpha(k_-^2-k^2))
\end{equation}
(the expression for ${\bf W}_{1,0}$ is exact).
Here it is denoted $k_\pm=((k\pm\delta_x)^2+\delta_y^2)^{1/2}$,
$s_\pm=(k\pm\delta_x)k_\pm^{-1}$ and $a_\pm=k_\pm^2+\pi^2$.
${\bf W}_{1,0}$ is a large-scale horizontal mode.

The paper is mainly concerned with the eigenvalue problem (\ref{eig})
in the three-dimensional invariant subspace, discussed above, where
the eigenfunction $\bf W$ and the operator $L$ take the forms (\ref{seriW}) and
(\ref{seriA}), respectively. The case of an unstable mode from this subspace
is called {\it small angle} instability, because the angles between
the wave vector $(k,0,\pi)$ of perturbed rolls and the wave vectors
$(k+\delta_x,\delta_y,\pi)$ of short-scale components of perturbation
(${\bf W}_{2,0}$ and ${\bf W}_{3,0}$ in the leading order) are
small. In section 5 we also study stability with respect to perturbations
of the form of rolls at an angle $\xi$ to the perturbed ones,
where $\xi$ is finite (and not asymptotically small). In the remaining
part of the section it is shown that any growing mode of $L$ belongs to one
of these two classes.

Consider the space ${\cal F}(\delta_x,\delta_y)$ of 4-component vector fields
of the form (\ref{fcom}), such that $\Wvel$ and $\Wtem$\\
$(i)$ are linear combinations of harmonics with wave vectors
$(m_1k+\delta_x,\delta_y,m_2\pi)$ or $(m_1k-\delta_x,-\delta_y,m_2\pi)$, where
$m_1$ and $m_2$ are integer,\\
$(ii)$ are symmetric about the vertical axis, i.e.\\
$$({\bf v}(x,y,z),\theta(x,y,z))=
(-v_x(-x,-y,z),-v_y(-x,-y,z),v_z(-x,-y,z),\theta(-x,-y,z)),$$
$(iii)$ satisfy the boundary conditions (\ref{bouc}),\\
$(iv)$ $\Wvel$ is divergence-free.\\
It was shown in OP2008 that $\cal F$ is $L$-invariant
for any (not necessarily small) $\delta_x$ and $\delta_y$.

Let the subspace $\tilde{\cal F}(\delta_x,\delta_y)$ be defined like
we have defined ${\cal F}(\delta_x,\delta_y)$, but omitting the condition
$(ii)$. Such subspaces also are $L$-invariant. We consider perturbations
which are doubly periodic on the $(x,y)$ plane. Let the domain of $L$ be
comprised of 4-component vector fields, for which $(iii)$ and $(iv)$ are
satisfied, and which have the same double periodicity on the $(x,y)$ plane.
The domain can be split into a sum of invariant spaces
$\tilde{\cal F}(\delta_x,\delta_y)$; hence we can assume that a mode belongs
to such an invariant subspace. Any mode can be represented as a sum
of a symmetric and an antisymmetric vector field; each of these vector fields
is itself a mode, since the subspaces of symmetric and an antisymmetric vector
fields are $L$-invariant. Moreover, in a coordinate system with the origin
shifted by half a period, $\pi/\delta_y$, in the $y-$direction (this shift does
not affect rolls, since they are independent of the $y$ coordinate) the
antisymmetric modes become symmetric.
In the case $\delta_y=0$, if integer $m$ and $n$ such that
$\delta_x/k=(2m+1)/(2n)$ exist, the shift of the origin by
$l=2\pi n/k=\pi(3m+1)/\delta_x$ along the $x-$direction turns a symmetric mode
into an antisymmetric one. For a given $k$ and $\delta_x$ the ratio
$(2m+1)/(2n)$ can be arbitrary close to $\delta_x/k$. Consequently, without any
loss of generality we consider henceforth only modes belonging to $\cal F$.

Eigenvalues of (\ref{seriL}) are perturbations of the ones of $L_0$.
Positive or slightly negative eigenvalues of $L_0$ are
associated with eigenvectors, whose wave vectors are either
$(\delta_x,\delta_y,0)$ (large-scale mode) or $(k\cos\xi,k\sin\xi,\pi)$
with $k$ close to $k_s$ (see, e.g., discussion in
Bernoff 1994). The latter eigenvectors are (\ref{v1}) rotated by the
angle $\xi$ about a vertical axis; we denote them by ${\bf U}(k,\xi)$.
Consider an eigenvector $\bf W$ of $L$
which belongs to some $\cal F$. The vector field
${\bf W}_0=\lim_{\varepsilon\to 0}\bf W$ can be one of the following:\\
(a) ${\bf W}_0=a_1{\bf W}_{1,0}+a_2{\bf W}_{2,0}+a_3{\bf W}_{3,0}$ (with
at least one $a_j\ne 0$ and ${\bf W}_{j,0}$ defined by (\ref{W01})-(\ref{W03})~);\\
(b) ${\bf W}_0={\bf U}(k,\xi)$ with $k$ close to $k_s$, $\xi=\rO(1)$ and
$\xi\pm2\pi/3=\rO(1)$;\\
(c) ${\bf W}_0=a_1{\bf U}(k,\xi_1)+a_2{\bf U}(k,\xi_2)$ with $k$ close to $k_s$,
$\xi_1-2\pi/3=\ro(1)$ and $\xi_2+2\pi/3=\ro(1)$.\\
In case (a) small-angle instability takes place studied in section 4,
in cases (b) and (c) finite angle instability considered in section 5.
Therefore, we examine all types of possibly growing perturbations.

\section{Stability of rolls: analytical results}

It is shown in Appendix B that in order to study stability of rolls
with respect to perturbations from the subspace constructed above,
it suffices to check, whether (for a given $P$, $\varepsilon$
and $\alpha$) there exist such $\delta_x$ and $\delta_y$ that
the following inequalities are satisfied:
\begin{equation}
\det{\cal A}>0
\label{zeroe}\end{equation}
or
\begin{equation}
S({\cal A})\tr{\cal A}-\det{\cal A}>0
\label{ime}\end{equation}
where
$$S({\cal A})=A_{11}A_{22}-A_{12}A_{21}+A_{11}A_{33}-A_{13}A_{31}+
A_{22}A_{33}-A_{23}A_{32}$$
is the sum of the three second order minors. The matrix $\cal A$
has an eigenvalue with a positive real part, if and only if at least one
of the inequalities is satisfied for some $\delta$'s.

It is calculated in Appendix A, that in the leading order
\begin{equation}
\det{\cal A}=D_0+\varepsilon^2D_1+\varepsilon^4D_2+\alpha D_3+
\alpha^2D_4+\alpha\varepsilon^2 D_5,
\label{det0}\end{equation}
where
$$\renewcommand{\arraystretch}{1.6}
\begin{array}{l}
D_0=
d_{01}\delta_x^6+d_{02}\delta_x^4\delta_y^2+d_{03}\delta_x^2\delta_y^6+
d_{04}\delta_y^{10},\\

D_1=
(\delta_x^2+\delta_y^2)^{-1}(d_{11}\delta_x^6+d_{12}\delta_x^4\delta_y^2+
d_{13}\delta_x^2\delta_y^4+d_{14}\delta_y^8),\\

D_2=
(\delta_x^2+\delta_y^2)^{-1}(d_{21}\delta_x^2\delta_y^2+d_{22}\delta_y^4),\\

D_3=d_{31}\delta_x^2\delta_y^4+d_{32}\delta_y^8,\\

D_4=
(\delta_x^2+\delta_y^2)(d_{41}\delta_x^2+d_{42}\delta_y^4),\\

D_5=
d_{51}\delta_x^2\delta_y^2(\delta_x^2+\delta_y^2)^{-1}+d_{52}\delta_y^4;
\end{array}
$$

\begin{equation}
S({\cal A})\tr{\cal A}-\det{\cal A}=E_0+
\varepsilon^2E_1+\varepsilon^4E_2+\alpha^2E_3+\alpha\varepsilon^2 E_4,
\label{min0}\end{equation}
where

$$\renewcommand{\arraystretch}{1.6}
\begin{array}{l}
E_0=
e_{01}\delta_x^6+e_{02}\delta_x^4\delta_y^2+e_{03}\delta_x^2\delta_y^4+
e_{04}\delta_y^{8},\\

E_1=
(\delta_x^2+\delta_y^2)^{-1}(e_{11}\delta_x^6+e_{12}2\delta_x^4\delta_y^2+
e_{13}\delta_x^2\delta_y^4+e_{14}\delta_y^6),\\

E_2=
(\delta_x^2+\delta_y^2)^{-1}(e_{21}\delta_x^4+e_{22}\delta_x^2\delta_y^2+e_{23}\delta_y^4),\\

E_3= e_3\delta_x^4\\

E_4= e_4\delta_x^2\delta_y^2(\delta_x^2+\delta_y^2)^{-1}.
\end{array}
$$

The values of $d_{ij}$ and $e_{ij}$ are given by
expressions (\ref{ddd}) and (\ref{eee}) in Appendix A.

To investigate stability of rolls we consider exhaustively different asymptotic
relations between $\alpha$ and $\varepsilon$ and different signs of $\alpha$.
We also consider two limit values of the Prandtl number.

Examples of areas on the $(k,R)$ plane where rolls are stable are shown on
Fig.~1 for several values of $P$. The area of stable rolls found numerically
(see section 8) is shaded; the instability boundaries determined analytically
are shown by lines. We use the standard notation (Busse and Bolton 1984;
Bernoff 1994; Mielke 1997; Getling 1998) for the instability modes and
respective instability boundaries. For a skew-varicose (SV) mode the associated
eigenvalue is real, and for an oscillatory skew-varicose (OSV) mode
the associated eigenvalues are complex; both modes exist for
$\delta_x\sim\delta_y$. A zigzag (ZZ) mode emerges for $\delta_x=0$ and
the associated eigenvalue is real. The mode responsible for the instability
for $P<0.782$ exists for $\delta_y\gg\delta_x\ne 0$ and the associated
eigenvalue is real; it is also called a skew-varicose mode (see, e.g., Mielke
1997). To distinguish this skew-varicose mode from the SV mode, we label
the former SV2.

\subsection{The case $\alpha^2\ll\varepsilon^4$}

Let $\delta_x^2\gg\delta_y^2$ and $\delta_x^2\ll\varepsilon^2$.
Then in the leading order
\begin{equation}
\det{\cal A}=\varepsilon^4 d_{21}\delta_y^2.
\label{det1}\end{equation}
$d_{21}$ is positive for all $P$ (see (\ref{ddd})\,), hence $\det{\cal A}>0$
as well, implying that for the assumed $\delta$'s the matrix has a
positive real eigenvalue and the rolls suffer from monotonous instability
(it can be shown that oscillatory instability does not emerge in this case).

\subsection{The case $\alpha>0$}

Assume again $\delta_x^2\gg\delta_y^2$ and $\delta_x^2\ll\varepsilon^2$.
In the leading order the determinant now includes $\alpha-$dependent terms:
$$\det{\cal A}=\varepsilon^4d_{21}\delta_y^2
+\alpha D_3+\alpha^2D_4+\alpha\varepsilon^2 D_5.$$
However, for $\alpha>0$ the terms involving $\alpha$ are positive.
The first term is positive, as discussed in the previous subsection,
hence rolls are also monotonously unstable in this case.

\subsection{The case $\alpha<0$, $P<P_1\approx 0.782$, $\alpha^2\sim\varepsilon^4$}

Denote by $P_1$ the Prandtl number which is the solution to the equation
$$d_{13}=2(d_{02}d_{22})^{1/2},$$
i.e.
\begin{equation}
-P^2+2P+2=2(2P^2(P+1))^{1/2}.
\label{P1}\end{equation}
The solution is
$$P_1=(3-2\sqrt2)(1+\sqrt{7+4\sqrt2})\approx0.782$$
(cf. Zippelius and Siggia 1982).
If $\delta_x^2\ll\delta_y^2$, $\delta_x^2\gg\delta_y^4$ and $\alpha^2\ll\varepsilon^2$,
the sum of asymptotically largest terms in (\ref{det0}) is
\begin{equation}
d_{02}\delta_x^4\delta_y^2+\varepsilon^2d_{13}\delta_x^2\delta_y^2+
\varepsilon^4d_{22}\delta_y^2,
\label{P782}\end{equation}
which is positive for $P<P_1$. Hence, for $P<P_1$ and the assumed $\alpha$'s
rolls are unstable.

\subsection{The case $\alpha<0$, $P>P_1$, $\alpha^2\sim\varepsilon^4$}

Re-write (\ref{det0}) as $\det{\cal A}={\cal D}_1+{\cal D}_2$, where
\begin{equation}
{\cal D}_1=D_0+\varepsilon^2D_1+\varepsilon^4d_{22}\delta_y^4(\delta_x^2+\delta_y^2)^{-1}
+\alpha D_3+\alpha^2D_4+\alpha\varepsilon^2d_{52}\delta_y^4
\label{trm1}\end{equation}
\begin{equation}
{\cal D}_2=\varepsilon^2
(\varepsilon^2d_{21}+\alpha d_{51})\delta_x^2\delta_y^2(\delta_x^2+\delta_y^2)^{-1}.
\label{trm2}\end{equation}
As proved in Appendix C, ${\cal D}_1<0$ for $P>P_1$ and the assumed $\alpha$'s.
${\cal D}_2>0$ if
\begin{equation}
{\rm SV:}\quad
\varepsilon^2>-\alpha f_1,\quad f_1={d_{51}\over d_{21}}={108\over 7}\pi^2k.
\label{f1}\end{equation}
For $\varepsilon^2\gg\delta_x^2\gg\delta_y^2$
${\cal D}_2$ is asymptotically larger than ${\cal D}_1$, hence (\ref{f1})
yields a boundary for monotonous instability of rolls.

Represent (\ref{min0}) as
$S({\cal A})\tr{\cal A}-\det{\cal A}={\cal E}_1+{\cal E}_2$, where
\begin{equation}
{\cal E}_1=E_0+\varepsilon^2E_1+\alpha^2E_3
\label{oi1}\end{equation}
\begin{equation}
{\cal E}_2=\varepsilon^2(\varepsilon^2E_2+\alpha E_4).
\label{oi2}\end{equation}
${\cal E}_1<0$ for $P>P_1$ and assumed $\alpha$'s.
${\cal E}_2>0$ if
$$e_{22}+e_4{\alpha\over\varepsilon^2}>2(e_{21}e_{23})^{1/2},$$
i.e.
\begin{equation}
{\rm OSV:}\quad
\varepsilon^2<-\alpha f_2,\quad f_2=-{e_4\over2(e_{21}e_{23})^{1/2}-e_{22}}=
\label{f2}\end{equation}
$${108(P+1)^2\pi^2k\over(P+3)(3P^2+2P+2)+3P^2(P+1)^{1/2}(P+5)^{1/2}}.$$
For $\varepsilon^2\gg\delta_x^2$ and $\varepsilon^2\gg\delta_y^2$,
${\cal E}_2$ is asymptotically larger than ${\cal E}_1$, hence (\ref{f2}) is
a condition for instability.

\subsection{The case $\alpha<0$, $\alpha^2\gg\varepsilon^4$}

Assuming, as above, $\varepsilon^2\gg\delta_x^2$ and $\varepsilon^2\gg\delta_y^2$,
we find that $\varepsilon^2\alpha E_4>0$ is
asymptotically the largest term in (\ref{oi2}). Hence, rolls are unstable.

\subsection{The case $\alpha<0$, $\alpha^2\sim\varepsilon^2$ and large $P$}

For $P\to\infty$, $f_1$ in (\ref{f1}) has a finite limit, and $f_2$
in (\ref{f2}) vanishes. Hence, for $P$ sufficiently large
the instability under the condition (\ref{f2}) can compete with the instability
occurring for $\alpha\sim\varepsilon$.
As shown in Appendix D, for large $P$, $\det{\cal A}>0$, if
\begin{equation}
{\rm ZZ:}\quad
\varepsilon^2<f_3\alpha^2,\quad f_3={9\pi^2P^2\over2(P+1)}.
\label{ipn}\end{equation}

\subsection{The case $\alpha<0$,
$\alpha^2\sim\varepsilon^4$ and $P$ slightly smaller than $P_1$}

Mielke (1997) showed that for $P<P_1$ rolls near the onset are always unstable.
More precisely, the following has been proved: for such $P$ there exists
a neighbourhood of the point $(k_s,R^s)$ in the $(k,R)$ plane, such that for
a given $R$ rolls of horizontal wave number $k$ (where they exist) are unstable.
The question, how the area where rolls are stable is modified, as $P$ becomes
smaller than $P_1$, has not been addressed in literature. We show below that
the area of stable rolls does not disappear abruptly and it still exists near
the onset, but its boundary does not include the point $(k_s,R^s)$ (cf.
figs.~1a,b and figs.~1c-f). As $P$ decreases, the area of stable rolls moves away
from this point, because the SV2 boundary moves to the left.

In the search of the horizontal scale ratios for which (\ref{det0}) is positive
for $P<P_1$, it was assumed in section 4.3
that $\delta_y^2\gg\delta_x^2\gg\delta_y^4$ and then in the leading order
$\det\cal A$ is given by (\ref{P782}). For $P$ slightly smaller than $P_1$
the expression (\ref{P782}) can be of the same order as
other terms in (\ref{det0}) not far from the onset. Consider a new small parameter
$\beta=P_1-P$. The maximum of (\ref{P782}) is admitted for
\begin{equation}
\delta_x^2=-{d_{13}\varepsilon^2\over2d_{02}}\equiv q\varepsilon^2,
\label{maxi}\end{equation}
and the maximum is equal to
$$-\varepsilon^4\delta_y^2{4(-(-P^2+2P+2)^2+8P^2(P+1))\over 81\pi^4P(P+1)^2}\approx
d_6\varepsilon^4\delta_y^2\beta,$$
where
$$d_6={4(-2(-2P_1+2)(-P_1^2+2P_1+2)+24P_1^2+16P_1)\over 81\pi^4P_1(P_1+1)^2}
\approx 5.0\cdot10^{-3}.$$

For $\delta_x$ defined by (\ref{maxi}) in the leading order the determinant is
\begin{equation}
\varepsilon^2(\varepsilon^2(\varepsilon^2(d_{01}+d_{02})q^3+\varepsilon^2d_{12}q^2
+\varepsilon^2d_{21}q+\alpha d_{51}q)+
\varepsilon^2d_6\delta_y^2\beta+(d_{03}q+d_{14})\delta_y^6)
\label{detsp}\end{equation}
which is a cubic polynomial of $\delta_y^2$. Its maximum is admitted, when
\begin{equation}
\delta_y^4=-{\varepsilon^2d_6\beta\over3(d_{03}q+d_{14})},
\label{maxiy}\end{equation}
and the maximum of (\ref{detsp}) is
$$\varepsilon^4s_3(\varepsilon^2s_1+\varepsilon\beta^{3/2}s_2+\alpha),$$
where
$$s_3=d_{51}q\approx0.0299,\ s_1=s_3^{-1}((d_{01}+d_{02})q^3+d_{12}q^2+d_{21}q)\approx0.00279,$$
$$s_2={2\over 3}d_6s_3^{-1}\biggl({-d_6\over 3(d_{03}q+d_{14})}\biggr)^{1/2}\approx0.0876.$$
The maximum is positive for
\begin{equation}
\alpha>-\varepsilon^2s_1-\varepsilon\beta^{3/2}s_2.
\label{f4}\end{equation}

However, the boundary of the SV2 instability defined by (\ref{f4}) turns out
to be in a poor agreement with the numerical results discussed in section 8.
For $P=0.6$ and $P=0.7$ (figs.~1a,b) the right boundary (\ref{f4}) of the area
of stable rolls is shifted far to the left compared to the computed one.
The SV2 boundary defined by the condition
$\max_{\delta_x,\delta_y}\det{\cal A}=0$ with all leading terms
in (\ref{det0a}) retained is still shifted too far to the left. The asymptotics
fails because the values (\ref{maxi}) and (\ref{maxiy}) of $\delta_x$ and
$\delta_y$, respectively, are of the order of 0.1 for the considered
overcriticalities $\varepsilon^2\sim 1$, while the asymptotic analysis is
applicable for infinitesimally small $\delta_x$ and $\delta_y$.
(For example, as we have found numerically, for $P=0.7$ the intersection
of the SV and SV2 boundaries is at $\alpha=-0.00172$ and $\varepsilon^2=0.583$.
The respective values (\ref{maxi}) and (\ref{maxiy}) are $\delta_x=0.126$ and
$\delta_y=0.413$. For the SV2 dominant mode on the stability boundary the
computed values are also large, $\delta_x=0.124$ and $\delta_y=0.332$, which
surprisingly do not differ much from the values obtained analytically.)

Consequently, we follow an alternative approach and assume that the SV2
instability boundary can be described as an equation, where it suffices to
retain two first non-vanishing terms in the Taylor expansion in $\varepsilon$
and $\beta$. The condition for the instability thus takes the form
\begin{equation}
{\rm SV2}:\quad
\alpha>\varepsilon^2h_1+\varepsilon\beta h_2,
\label{f4n}\end{equation}
for some coefficients $h_1$ and $h_2$, which can be determined numerically.
The intersections of the SV2 boundary with the SV and OSV boundaries have been
computed for $P=0.6$ and $P=0.7$ (see figs.~1a,b). The minimum (over the
coefficients $h_i$) of the maximum (over the four points of intersection)
relative error is equal to 0.14, it is admitted for
\begin{equation}
h_1=-0.0012\hbox{ and }h_2=-0.018\,.
\label{f4nn}\end{equation}
By the relative error we understand the ratio $|(\alpha_c-\alpha_t)/\alpha_c|$,
where $\alpha_c$ is the computed value (see section 8) and $\alpha_t$ is found
from (\ref{f4n}) at the points of intersection.
We have also computed several points on the SV2 boundary in the regions of
other instabilities of rolls, employing the fact that the
respective (local) maximum of $\lambda$ over $\delta_x$ and $\delta_y$
is admitted for $\delta_y$ much larger than for the other instabilities.
The computed values agree well with (\ref{f4n}), (\ref{f4nn}) (see figs.~1a,b).
Fitting of $s_1$ and $s_2$ in (\ref{f4}) yields a much higher (about 0.5)
minimum of the maximum over the four points relative error.

\section{The finite angle case}

To analyse stability of rolls of horizontal wave number $k$ to rolls
of wave number $k_p$, which are rotated by angle $\xi$ with finite $\xi$, we
use center manifold reduction with the center eigenspace
spanned by eigenvectors (\ref{v1}) with wave vectors
$(k,0,\pi)$ and $(k_p\cos\xi,k_p\sin\xi,\pi)$. We perform the
reduction like in (Podvigina and Ashwin 2007).
Here only results of calculations are presented.
Periodicity in horizontal directions of the considered rolls implies that
periodicity cells are parallelograms (and not squares as {\it ibid}).

Restricted to the two-dimensional (${\bf C}^2$) center manifold,
the system has the form
\begin{equation}
\label{ds}
\renewcommand{\arraystretch}{1.6}
\begin{array}{l}
\dot z_1=\lambda_1z_1+z_1(A_1|z_1|^2+A_2|z_2|^2),\\
\dot z_2=\lambda_2z_2+z_2(A_3|z_1|^2+A_4|z_2|^2),
\end{array}
\end{equation}
where $z_1$ and $z_2$ are coordinates in the center manifold, along the
directions $(k,0,\pi)$ and $(k_p\cos\xi,k_p\sin\xi,\pi)$, respectively.
The reduction is performed for $R=R_c(k)$.
We are interested in $k_p$ close to $k$ (otherwise $\lambda_2$ is of the order
of one and negative and thus the rolls $(z_1,0)$ are stable near the onset).

For $\varepsilon$ defined by (\ref{r2c}) and $k_p$ close to $k$,
the coefficients of linear terms in (\ref{ds}) are
$$
\renewcommand{\arraystretch}{1.6}
\begin{array}{l}
\lambda_1=P(P+1)^{-1}k^2a^{-2}\varepsilon^2,\\
\lambda_2=
P(P+1)^{-1}\biggl(k^2a^{-2}\varepsilon^2-4(k_p-k)^2-8\alpha(k_p-k)\biggr).
\end{array}
$$
For small $k_p-k$ the differences $A_1-A_4$ and $A_2-A_3$ are small,
coefficients of cubic terms in (\ref{ds}) are
$$A_1\sim A_4=-{0.125P\over(P+1)},$$
$$A_3\sim A_2=-{0.125P\over(P+1)}-{a(1-\cos^2\xi)\over3(P+1)}
\biggl((1-\cos\xi){2aq_++2Pa^2\over P\Delta_+}+
(1+\cos\xi){2aq_-+2Pa^2\over P\Delta_-}\biggr)$$
$$-{\pi^2\over2(P+1)}\biggl((1-\cos\xi)^2{Pq_++3h_+a\over \Delta_+}+
(1+\cos\xi)^2{Pq_-+3h_-a\over \Delta_-}\biggr),$$
where
$$h_\pm=2k^2(1\pm\cos\xi),\ q_\pm=4\pi^2+h_\pm,\ \Delta_\pm=q_\pm^3-Rh_\pm.$$

The amplitude of emerging rolls is
$$|z_1|^2=-\lambda_1/A_1,$$
and four eigenvalues of (\ref{ds}) linearised around the steady state are
\begin{equation}
\label{eig4}
-2\lambda_1,\quad \lambda_2+A_3|z_1|^2,\quad 0,\quad 0,
\end{equation}
hence the instability condition is
$$\varepsilon^2k^2a^{-2}(1-A_3^{max}/A_1)-4(k-k_p)^2-8\alpha(k-k_p)>0,$$
i.e.~instability occurs if
$$\varepsilon^2<f_5\alpha^2,\quad f_5=-{4A_1a^2\over k^2(A_1-A_3^{max})},$$
where by $A_3^{max}$ we have denoted the maximum of $A_3$ in $\xi$.

For a finite $P$ the instability boundary is below the boundary
defined by (\ref{f2}). For large $P$, the limits of $A_1$ and $A_3$ are finite,
hence $f_5<f_3$, and finite-angle instabilities do not affect the area
of stability of rolls. The instability with respect to rolls rotated by
$\xi=\pi/2$ is called the cross-roll instability. Note that
the maximum of $A_3$ can be admitted for a $\xi\ne\pi/2$,
but we do not consider here the problem of maximisation of $A_3$ in $\xi$.

For $\xi$ close to $2\pi/3$ the center eigenspace also involves
rolls with the direction of the axes rotated by $-2\pi/3$. The system
restricted to the three-dimensional (${\bf C}^3$) center manifold is
\begin{equation}
\label{ds3}
\renewcommand{\arraystretch}{1.6}
\begin{array}{l}
\dot z_1=\lambda_1z_1+z_1(A_1|z_1|^2+A_2|z_2|^2+B_1|z_3|^2),\\
\dot z_2=\lambda_2z_2+z_2(A_3|z_1|^2+A_4|z_2|^2+B_2|z_3|^2),\\
\dot z_3=\lambda_2z_3+z_3(A_5|z_1|^2+A_6|z_2|^2+B_3|z_3|^2).
\end{array}
\end{equation}
However, the eigenvalues determining stability of rolls are
(\ref{eig4}), examined above.

\section{Growth rates}
\label{gro}

In this section we find orders of growth rates of the dominant unstable modes.
If entries of the matrix $\cal A$ are of different orders, it is possible
to calculate dominant eigenvalues, like it was in the case of rotating layer
in OP2008. In the present problem often almost all entries of $\cal A$
turn out to have the same asymptotics, and only orders of growth rates
can be determined. Also we find orders of coefficients $a_j$, $j=$1,2,3,
of the most unstable mode
\begin{equation}
\label{emode}
{\bf W}=a_1\widetilde{\bf W}_1+a_2\widetilde{\bf W}_2+a_3\widetilde{\bf W}_3.
\end{equation}

Unstable modes (or instabilities) can be roughly categorised into five
different types\footnote{{\bf E-l} stands for Eckhaus-like instability.
Maximisation of the growth rate in $\delta_x$ and $\delta_y$
yields only the horizontal wave number of the most unstable mode, see OP2008
and section 6.5. The conventional Eckhaus instability is a particular case
of the {\bf E-l} instability for $\delta_y=0$.}:
$$
\renewcommand{\arraystretch}{1.6}
\begin{array}{ll}
{\rm SV:}&\delta_x\sim\delta_y\\
{\rm SV2:}&\delta_x\ll\delta_y\\
{\rm OSV:}&\delta_x\sim\delta_y\\
{\rm ZZ:}&\delta_x=0\\
{\rm E-l:}
&2k\delta_x+\delta_y^2=-2k\alpha
\end{array}
$$

Similarly to section 4, we consider different asymptotic relations between
$\alpha$ and $\varepsilon$. Our findings are summarised in Table 1, where
$$\xi_1={(P^2-2P-2)^2-8P^2(P+1)\over 9\pi^2(P+1)^2P},\
\xi_2={4(2k)^{1/2}\over 3\pi(P+1)^{1/2}},\
\xi_3={4P\over(P+1)},\ \xi_4={8\over9\pi^2P}.$$
\pagebreak

\vspace*{10mm}
\noindent
Table 1.
Possibly dominant instability modes for various asymptotic relations between
$\alpha$ and $\varepsilon$ and values of $P$. The last column presents
eigenvalues, when they can be calculated, or their orders of magnitude
otherwise. (Hence, often it remains unclear which mode is dominant.)

\medskip\hspace*{-23mm}
\renewcommand{\arraystretch}{1.4}
\begin{tabular}{|c|c|c|c|c|}\hline
Relations&Conditions for&Type of&$\delta_x$ and $\delta_y$&
Eigenvalues\\
between $\alpha$ and $\varepsilon$&existence&the mode&&\\\hline
$\alpha^2\ll\varepsilon^4$&none& SV &$\delta_x\sim\delta_y\sim\varepsilon$&
$\lambda\sim\varepsilon^2$\\
& $P<P_1$& SV2 &$\delta_x\ll\delta_y,\ \delta_x\sim\varepsilon$&
$\lambda=\xi_1\varepsilon^2$\\\hline
$\alpha^2\sim\varepsilon^4$&$\varepsilon^2>-f_1\alpha$&
 SV &$\delta_x\sim\delta_y\sim\varepsilon$&$\lambda\sim\varepsilon^2$\\
&$P<P_1$& SV2 &$\delta_x\ll\delta_y,\ \delta_x\sim\varepsilon$&
$\lambda=\xi_1\varepsilon^2$\\
&$\varepsilon^2<-f_2\alpha$& OSV &$\delta_x\sim\delta_y\sim\varepsilon$&
Re$(\lambda)\sim\varepsilon^2$\\\hline
$\varepsilon^4\ll\alpha^2\ll\varepsilon^{4/3}$&$\alpha>0$&
 SV &$\delta_x^2\sim\delta_y^2\ll\varepsilon\alpha^{1/2}$&
$\lambda=\xi_2\varepsilon\alpha^{1/2}$\\
{\rm or}&$\alpha<0$, &
 OSV &$\delta_x^2\sim\delta_y^2\sim\varepsilon\alpha^{1/2}$&
Re$(\lambda)\sim\varepsilon\alpha^{1/2}$\\
$\alpha^2\sim\varepsilon^{4/3}$&$\varepsilon^2<-f_2\alpha$&&&\\\hline
$\alpha^2\sim\varepsilon^2$&$\alpha<0,$& ZZ &
$\delta_x=0,\ \delta_y\sim\varepsilon$&$
\lambda=-\xi_4\varepsilon^2+\xi_3\alpha^2$\\
&$\varepsilon^2>-f_2\alpha,$&&&\\
&$\varepsilon^2<f_3\alpha^2$&&&\\\hline
$\alpha^2\gg\varepsilon^{4/3}$&none& E-l &$2k\delta_x+\delta_y^2=-2k\alpha$&
$\lambda=\xi_3\alpha^2$\\\hline
\end{tabular}

\vspace*{10mm}
\subsection{The case $\alpha^2\ll\varepsilon^4$}

In this case there exists a growing mode
$${\rm SV: }\quad\lambda\sim\varepsilon^2,\hbox{ for }
\delta_x^2\sim\delta_y^2\sim\varepsilon^2,$$
since it can be easily shown that $\det{\cal A}>0$ for some
\begin{equation}
\delta_x^2\sim\delta_y^2\sim\varepsilon^2.
\label{ass}\end{equation}
If (\ref{ass}) holds, $\tr{\cal A}\sim\varepsilon^2$, $S{\cal A}\sim\varepsilon^4$
and $\det{\cal A}\sim\varepsilon^6$, implying that eigenvalues are $\sim\varepsilon^2$.
For the assumed dependence of $\delta$'s on $\varepsilon$, after the change of
variables $\widetilde{\bf W}_2\to\varepsilon\widetilde{\bf W}_2$ all coefficients (except for
$\widetilde A_{32}$) become of the same order in $\varepsilon$, implying that for
the associated eigenmode\quad $a_1/a_2\sim a_3/a_2\sim\varepsilon$.

For $P<P_1$ the expression (\ref{det0a}) can be positive also, if
$\delta_y^2\gg\delta_x^2\sim\varepsilon^2$. For $\delta_x^2+\delta_y^2\gg\varepsilon$,
the matrix (\ref{atil}) has an eigenvalue close to
$-P(\delta_x^2+\delta_y^2)$. In the leading order the associated eigenvector
is $\widetilde{\bf W}_1+\xi_2\widetilde{\bf W}_2+\xi_3\widetilde{\bf W}_3$,
where $\xi_2=\widetilde A_{21}/(\widetilde A_{11}-\widetilde A_{22})$ and
$\xi_3=\widetilde A_{31}/(\widetilde A_{11}-\widetilde A_{33})$.
Two remaining eigenvalues are eigenvalues of the matrix
\begin{equation}
\label{matr2}
\left[
\begin{array}{cc}
\widetilde A_{22}-\xi_2\widetilde A_{12}&\widetilde A_{23}-\xi_2\widetilde A_{13}\\
\widetilde A_{32}-\xi_3\widetilde A_{12}&\widetilde A_{33}-\xi_3\widetilde A_{13}
\end{array}
\right].
\end{equation}
In the leading order they are
\begin{equation}
\label{lacr}
\lambda=-C_4(4k^2\delta_x^2+\delta_y^4+4\alpha k\delta_y^2)-\varepsilon^2C_3+
\varepsilon^2C_5\delta_y^2{(3\delta_x^2-\delta_y^2)\over(\delta_x^2+\delta_y^2)^2}
\end{equation}
$$
\pm\Biggl[\varepsilon^4
\Bigl(C_3+C_5\delta_y^2{(3\delta_x^2-\delta_y^2)\over(\delta_x^2+\delta_y^2)^2}\Bigr)^2+
4C_4k^2\delta_x^2(\delta_y^2+2\alpha k)\Bigl(\varepsilon^2
C_5{\delta_y^2\over(\delta_x^2+\delta_y^2)^2}-
4C_4(\delta_y^2+2\alpha k)\Bigr)\Biggr]^{1/2},
$$
where
$$C_5={b^2\pi^2\over2k^2P}.$$

Calculating their maxima in $\delta_x$ and $\delta_y$ we find the most unstable mode:
$${\rm SV2: }\quad\lambda=\varepsilon^2
{(P^2-2P-2)^2-8P^2(P+1)\over 9\pi^2(P+1)^2P},\hbox{ for }$$
$$\delta_x^2=\varepsilon^2{8(P+1)^2-(P^2-2P-2)^2\over 18(P+1)P^2},
\ \delta_y\gg\delta_x.$$
These relations between $\delta$'s, $\varepsilon$ and $\lambda$ imply that
the associated eigenvector $(a_1,a_2,a_3)$ of the matrix $\widetilde{\cal A}$
(\ref{atil}) has components with the asymptotics
$a_1/a_2\sim\varepsilon^2\delta_y^{-1}$ and $a_3/a_2\sim\varepsilon$.

\subsection{The case $\alpha^2\sim\varepsilon^4$}

Dominant eigenvalues and asymptotic relations between the coefficients
$a_1,a_2,a_3$ for the eigenmodes SV and SV2
are the same as above. The SV mode is growing
if (\ref{f1}) holds, and SV2 if $P<P_1$.

A growing oscillatory mode can exist, if (\ref{f2}) holds true. As discussed
in Appendix B, condition (\ref{ime}) does not guarantee its existence. However,
if such mode exists for all $\alpha^2\gg\varepsilon^4$, $\alpha<0$, (see
section 6.4), by continuity it exists for some $\alpha^2\sim\varepsilon^4$.
The maximal growth rate of the mode
$${\rm OSV:}\quad\hbox{Re}(\lambda)\sim\varepsilon^2$$
is admitted for
$$\delta_x^2\sim\delta_y^2\sim\varepsilon^2.$$
By the same arguments as for the SV mode, for the OSV eigenmode
with the maximal growth rate and $a_1/a_2\sim a_3/a_2\sim\varepsilon$.

\subsection{The case $\varepsilon^{4/3}\gg\alpha^2\gg\varepsilon^4$, $\alpha>0$}

We employ the same change of variables as in section 6.1, and maximisation
in $\delta_x$ and $\delta_y$ yields the maximal growth rate for SV modes
\begin{equation}
{\rm SV:}\quad\lambda={1\over 2P^{1/2}}d_{51}^{1/2}\varepsilon\alpha^{1/2}=
{4(2k)^{1/2}\over 3\pi(P+1)^{1/2}}\varepsilon\alpha^{1/2},\hbox{ for }
\delta_x^2=\delta_y^2\ll\varepsilon\alpha^{1/2}.
\label{type1}\end{equation}
The eigenmode coefficients have the asymptotics
$a_1/a_2\sim\varepsilon\alpha^{-1/2}$ and $a_3/a_2\sim\varepsilon^2(\alpha^{1/2}\delta_x)^{-1}$.
The SV2 mode has the growth rate $\rO(\varepsilon^2,\alpha^2)$,
which is asymptotically smaller than (\ref{type1}).

\subsection{The case $\varepsilon^{4/3}\gg\alpha^2\gg\varepsilon^4$, $\alpha<0$}

For the dominant oscillatory mode the maximal growth rate and
$(\delta_x,\delta_y)$ for which it is admitted are:
\begin{equation}
{\rm OSV:}\quad\hbox{Re}(\lambda)\sim\varepsilon\alpha^{1/2},
\hbox{ at }\delta_x^2\sim\delta_y^2\sim\varepsilon\alpha^{1/2}.
\label{type3}\end{equation}
This can be obtained by the following arguments. Assume
\begin{equation}
\delta_x^2\sim\delta_y^2\ll\varepsilon\alpha^{1/2}.
\label{assu}\end{equation}
Consider the cubic equation $\det(\lambda{\rm\bf I}-{\cal A})=0$. The assumption
(\ref{assu}) implies, by virtue of the standard formulae for roots of cubic equations,
existence of complex roots with a positive Re$\lambda\sim(\det{\cal A})^{1/3}
\sim\varepsilon^2\alpha\delta_x^2\delta_y^2(\delta_x^2+\delta_y^2)^{-1}$.
Since this holds true for any $(\delta_x,\delta_y)$ satisfying (\ref{assu}),
this relation
remains true for a $\delta_x^2\sim\delta_y^2\sim\varepsilon\alpha^{1/2}$.
The last two asymptotic relations imply that the associated eigenvector
$(a_1,a_2,a_3)$ of the matrix $\widetilde{\cal A}$ (\ref{atil}) has components
with the asymptotics
$a_1/a_2\sim\alpha^{1/2}$ and $a_3/a_2\sim\alpha^{3/4}\varepsilon^{-1/2}$.
For the SV2 mode the growth rate is $\sim\varepsilon^2$ or
$\sim\alpha^2$, i.e. it is asymptotically smaller than (\ref{type3}).

\subsection{The case $\alpha^2\gg\varepsilon^{4/3}$}

Maximisation of (\ref{lacr}) in $\delta_x$ and $\delta_y$ yields
that the maximal growth rate
\begin{equation}
\label{type4}
{\rm E-l:}\quad\lambda=4C_4k^2\alpha^2
\end{equation}
is admitted for
$$(k_\pm^2-k^2)=-2\alpha k;$$
the associated eigenvectors are either ${\bf W}_2$ or ${\bf W}_3$.
(Note that growth rates are asymptotically smaller than (\ref{type4}),
unless $\delta_x^2+\delta_y^2\gg\varepsilon$; if this asymptotic relation
is satisfied, (\ref{lacr}) employed in maximisation is valid, see section 6.1.)
Alternatively, (\ref{type4}) can be obtained directly from (\ref{AA}).
For the OSV mode $\hbox{Re}(\lambda)\sim\varepsilon^{4/3}$,
which is asymptotically smaller, than (\ref{type4}).

\subsection{The case $\alpha^2\sim\varepsilon^{4/3}$}

The maximal growth rate is $\rO(\alpha^2)\sim\rO(\varepsilon^{4/3})$.
For $\alpha>0$ it is given by (\ref{type1}), if
$$\varepsilon^2>\alpha^3{9\pi^2P\over 2k(P+1)},$$
or by (\ref{type4}) otherwise. For $\alpha<0$ it is either (\ref{type3}),
or (\ref{type4}).

\subsection{The case $\alpha^2\sim\varepsilon^2$, $\alpha<0$ and large $P$}

As noted in section 4.7, in the limit $P\to\infty$ the ZZ instability
with $\delta_x=0$ competes with the OSV instability and becomes of
importance near the onset. If $\delta_x=0$ and $\delta_y\gg\varepsilon$, the
eigenvalues of the matrix $\cal A$ are
$$\lambda_1=-P\delta_y^2,\quad \lambda_2=-2\varepsilon^2C_3-C_4\xi,\quad
\lambda_3=-\varepsilon^2b^2{\pi^2\over4Pk^2}-C_4\xi,$$
where
$$\xi=\alpha k\delta_y^2+\delta_y^4/4.$$
For large $P$, $\lambda_3>\lambda_2$, maximisation of $\lambda_3$ in $\delta_y$
yields the maximal growth rate
$$\lambda_{\max}=-\varepsilon^2b^2{\pi^2\over4Pk^2}+4C_4k^2\alpha^2.$$
The associated eigenvector has asymptotics $a_1/a_3\sim1$ and $a_2=0$.

\section{Asymptotics of neglected terms in equations for stability boundaries}
\label{errors}

Expressions (\ref{f1}), (\ref{f2}) and (\ref{ipn}) determining
stability of rolls are only asymptotically correct. In this section
we estimate the asymptotic order of errors in calculation of boundaries,
relying on the known orders of the remainder terms in (\ref{det0a}) and (\ref{min0a}).

In the course of derivation of an equation defining the SV instability boundary,
$\det{\cal A}$ has been expressed in section 4.4 as a sum of ${\cal D}_1$
(\ref{trm1}) and ${\cal D}_2$ (\ref{trm2}), where ${\cal D}_1$ is negative
and involves terms
${\rm O}(\delta^6+\varepsilon^2\delta^4+\varepsilon^4\delta_y^4\delta^{-2})$
(here $\delta^2=\delta_x^2+\delta_y^2$\,) and ${\cal D}_2$ can be positive
and involves terms ${\rm O}(\varepsilon^4\delta_x^2\delta_y^2\delta^{-2})$.
The inequality (\ref{f1}) is a restatement of the condition ${\cal D}_2>0$.
Near the boundary $\varepsilon^2\gg\delta_x^2\gg\delta_y^2$
must be satisfied so that the sum (\ref{trm2}) were positive. Under this condition
${\cal D}_1$ is asymptotically smaller than ${\cal D}_2$ and hence
asymptotic corrections to ${\cal D}_1$ do not affect the boundary.
Upon reintroduction of the terms omitted in (\ref{det0a}), that are not
asymptotically smaller than ${\cal D}_1$, (\ref{trm2}) becomes
$${\cal D}_2=\varepsilon^2\biggl(\varepsilon^2(d_{21}+{\rm O}(\varepsilon^2,\alpha))+
\alpha(d_{51}+{\rm O}(\varepsilon^2,\alpha))\biggr)
\delta_x^2\delta_y^2(\delta_x^2+\delta_y^2)^{-1}$$
and thus the equation for the boundary takes the form
\begin{equation}
{\rm SV}:\quad
\varepsilon^2>-\alpha f_1+{\rm O}(\alpha^2).
\label{f1n}\end{equation}
Since at the boundary the factor in front of
$\delta_x^2\delta_y^2\delta^{-2}$ vanishes, it can be shown using this
analysis that the condition
$$\max_{\delta_x,\delta_y}\det{\cal A}=0,$$
defining the boundary, implies $\delta_x=\delta_y=0$.

The OSV instability boundary (\ref{f2}) has been found from the condition
that ${\cal E}_2$ (\ref{oi2}) vanishes. By the same arguments as above, near the boundary
$\varepsilon^2\gg\delta_x^2\sim\delta_y^2$ must be satisfied for the sum of
${\cal E}_1$ and ${\cal E}_2$ to be positive and hence asymptotic corrections to
${\cal E}_1$ again do not affect the boundary. With the omitted terms of
(\ref{min0a}) reintroduced, (\ref{oi2}) becomes
$${\cal E}_2=\varepsilon^2\biggl(\delta_x^2+\delta_y^2)^{-1}
(\varepsilon^2(e_{21}+{\rm O}(\varepsilon^2,\alpha))\delta_x^4+
(\varepsilon^2e_{22}+\alpha e_4+{\rm O}(\varepsilon^4,\alpha^2))\delta_x^2\delta_y^2+
\varepsilon^2(e_{23}+{\rm O}(\varepsilon^2,\alpha))\delta_y^4\biggr).$$
This expression results in the equation for the instability boundary in the form
\begin{equation}
{\rm OSV:}\quad
\varepsilon^2<-\alpha f_2+{\rm O}(\alpha^2).
\label{f2n}\end{equation}
Again, it can be shown that at the boundary $\delta_x=\delta_y=0$.

In Appendix D the ZZ instability boundary is calculated from the condition that
$$\max_{\delta_y}\det{\cal A}=0,$$
where $\det{\cal A}$ is given by (\ref{lp1}).
Since on the boundary $\alpha^2\sim\varepsilon^2$, the equation
with the omitted terms reintroduced takes the form
\begin{equation}
\max_{\delta_y}\biggl(-\delta_y^2(2\varepsilon^2C_3+C_4\xi)(\varepsilon^2b^2{\pi^2\over4k^2}+
PC_4\xi)+{\rm O}(\delta_y^{14},\alpha^2\delta_y^8,\alpha^6\delta_y^2)\biggr)=0.
\label{lp1n}\end{equation}
The maximum is attained for
$$\delta_y^2=2\alpha k+{\rm O}(\alpha^2)$$
and the condition for the instability is thus
$${\rm ZZ:}\quad\varepsilon^2<f_3\alpha^2+{\rm O}(\alpha^3).$$

\section{Stability of rolls: numerical results}
\label{numer}

To examine stability of rolls of wave number $k$, we solve numerically
(with an adapted version of the code of Zheligovsky 1993)
the problem (\ref{eig}) for the eigenfunction
\begin{equation}
\label{ww}
{\bf W}={\rm e}^{{\rm i}\delta_xx+{\rm i}\delta_yy}
\sum_{m=-M}^{m=M}\sum_{n=0}^{n=N}\left(
\begin{array}{c}
w_{mn}^1{\rm e}^{{\rm i}mkx}\cos\pi nz\\
w_{mn}^2{\rm e}^{{\rm i}mkx}\cos\pi nz\\
w_{mn}^3{\rm e}^{{\rm i}mkx}\sin\pi nz\\
w_{mn}^4{\rm e}^{{\rm i}mkx}\sin\pi nz
\end{array}
\right).
\end{equation}
In computations, the cut off of the series at $N=M=15$ suffices (the spectrum of
the solution in the Fourier space decays by at least 12 orders of magnitude).
Location of the maximum of ${\rm Re}(\lambda)$ in $\delta_x$ and $\delta_y$ has
been determined with the precision of $10^{-4}$ (or $2\cdot10^{-5}$, if
$\delta_x$ and $\delta_y$ are below $0.01$) which allows us to find correctly
at least two significant digits of $\lambda$.

The dominant eigenvalues of (\ref{eig}),(\ref{ww}) and the values of $\delta_x$
and $\delta_y$ where the maximum is admitted are shown on fig.~2 for $P=2$ and
$k=2.15$ (thin vertical line on fig.~1c). In the interval $658.5\le R\le 658.8$,
i.e. for a small overcriticality, the Eckhaus mode with $\delta_y=0$
is dominant, the values of $\delta_x$ and $\lambda$ are close to the ones given
in the Table (according to the Table, $\lambda=0.0136$ and $\delta_x=0.071$).
In the interval $658.9\le R\le670$ the dominant eigenvalues are complex,
they are associated with the OSV eigenmode. The change of the type of
the dominant mode implies a discontinuity of $\delta$'s. In the interval
$658.9\le R\le661$, ${\rm Re}(\lambda)$ depends linearly on
$\varepsilon=(R-R_c)^{1/2}$, in agreement with the Table (in fact, for smaller
$R$, where the instability is subdominant, this asymptotics for the eigenvalue
of the OSV mode was also confirmed numerically), as predicted in section 6.
For higher $R$ the dependence is different, because, as noted in section 7,
near the SV and OSV boundaries $\delta_x$ and $\delta_y$ become asymptotically
smaller than $\varepsilon$, while in section 6 we have assumed
$\delta_x={\rm O}(\varepsilon)$ and $\delta_y={\rm O}(\varepsilon)$.
In agreement with section 7, we observe that ${\rm Re}(\lambda)$,
$\delta_x$ and $\delta_y$ vanish at $R_{\rm OSV}$ and $R_{\rm SV}$, where
$R_{\rm OSV}$ and $R_{\rm SV}$ denote the critical values of $R$
for the OSV and SV instabilities. In the interval $(R_{\rm OSV},R_{\rm SV})$,
where rolls are stable, the maximal growth rate is zero, admitted for
$\delta_x=\delta_y=0$. For $R>R_{\rm SV}$ the SV mode is dominant.
Near $R_{\rm OSV}$ and $R_{\rm SV}$, ${\rm Re}(\lambda)\sim(R_{\rm OSV}-R)^2$
and ${\rm Re}(\lambda)\sim(R_{\rm SV}-R)^2$, respectively (while no power law
asymptotics has been found for $\delta_x$ and $\delta_y$, except for
$\delta_y$ is almost linear near $R_{\rm SV}$).
Consequently, the SV and OSV boundaries are found by linear extrapolation of
$({\rm Re}(\lambda))^{1/2}$ through two computed points close to the boundary.
Near the ZZ and SV2 boundaries $\lambda$ depends on $R$ linearly, and we find
the instability boundary by linear interpolation.

The areas of stable rolls found numerically are shown on fig.~1 for several
values of $P$. The difference between the SV, OSV and ZZ boundaries predicted
theoretically and found numerically agrees with the estimations of
the remainder terms obtained in section 7. For small $\alpha$ and $\varepsilon$,
the theoretical and numerical boundaries visually coincide, and the discrepancy
remains small on increasing $\alpha$. The area of stable rolls
found numerically is shifted up compared to the one determined analytically,
indicating that the contribution of the omitted terms is positive for OSV and
ZZ instabilities, and negative for the SV instability.

In view of the good agreement of the analytical and numerical
results for these three boundaries, the disagreement for the SV2
boundary is surprising. A possible explanation is that the omitted
in (\ref{det0a}) terms involving $\delta_x$ (which are of no importance
for the SV and ZZ instabilities for which $\delta_x=0$ -- e.g.,
$C\varepsilon^2\delta_x^4\delta_y^2$) can turn out to be relatively strongly
negative and come into play already at $\varepsilon=1$.
A more plausible explanation is that for the SV2 mode the employed asymptotic
expansions of the operator of linearisation $L$, its eigenvectors and
eigenvalues are valid for much smaller $\alpha$ and $\varepsilon$ than for
other instabilities, because the values of $\delta_x$ and $\delta_y$
at the SV2 boundary, maximising the eigenvalue, are relatively large.
Note, that the same asymptotic expansion was employed in other analytic studies
of the problem, cited in the Introduction.

\section{Conclusion}

We have presented a complete analytical study of stability of rolls
near the onset of convection to perturbations, which are doubly periodic in
horizontal directions. In all earlier studies only instabilities of rolls
to certain classes of perturbations were shown.

In pursuit of this goal, we have, first, shown that without any loss of
generality any instability mode is responsible for either the small-angle,
or finite-angle instability. Second, for the small-angle instability modes
we have derived inequalities determining regions of stability of rolls.
The problem involves four small parameters; while deriving the
instability conditions we have considered all asymptotic relations between
the small parameters. Finally, we have calculated boundaries for
the finite-angle instability; it turns out that consideration of finite-angle
instability modes does not modify the region of stability of rolls.

In our analysis only the asymptotically largest terms have been taken into
account. A question often arises, whether enough terms of asymptotic expansions
have been calculated at various intermediate stages. In (\ref{atil}) orders
of the omitted terms in the matrix are given, implying that
the omitted terms in expressions (\ref{det0a}) and (\ref{min0a}), used here
to analyse stability, are irrelevant sufficiently close to onset.

This small-angle instability of rolls was studied before, and the SV,
OSV and ZZ instability boundaries found here coincide with the earlier results.
Our novel results concerning the stability boundaries include the following
ones: We have examined the dependence of the SV2 boundary on $P$ for
$0.543<P<0.782$. For decreasing $P$, the boundary of the region of stable rolls
on the $(k,R)$ plane moves to the left away from the point $(k_s,R^s)$. We have
established the asymptotics of the maximum growth rates and the associated
eigenmodes (see (\ref{emode})\,) considering exhaustively different relations
between $\alpha$ and $\varepsilon$. We have derived asymptotic equations
describing the regions of the instabilities and estimated remainders
in these equations.

The approach that we have followed can be applied to study instabilities of
stripe patterns with respect to large-scale perturbations in a generic system,
where a large-scale neutral mode exists. Existence of the invariant subspace
relies only on the structure of equations of convection, where the linear part
preserves wave vectors and nonlinearity is of the second order. Equations
((\ref{w1}) and (\ref{w2})) defining the entries of matrix $\cal A$ are
general, they remain valid for any other system defined by arbitrary mappings
$L_j$. Stability is analysed by examining the inequalities
(\ref{zeroe}) and (\ref{ime}). This analysis is, perhaps, the most difficult
part. It may change significantly for other systems with different asymptotics
of the entries of the matrix $\cal A$, resulting in different asymptotics
involved in the inequalities defining instability regions.

\bigskip
{\bf Acknowledgements}

\bigskip
Part of the research was carried out during my visits
to the Observatoire de la C\^ote d'Azur (Nice, France) in September --
December 2007 and 2008. I am grateful to the French Ministry of Education
for financing my research visits to the Observatoire de la C\^ote d'Azur.
I was also partially supported by grants ANR-07-BLAN-0235
OTARIE from Agence nationale de la recherche (France) and
07-01-92217-CNRSL{\Large\_}a from the Russian foundation for basic research.

\appendix
\section{Calculation of the matrix $\cal A$}

In this Appendix we calculate in the leading order the entries of the matrix
$\cal A$ of the restriction of $L$ on the invariant subspace spanned by
${\bf W}_j$, $j=1,2,3$, and expressions for $\det\cal A$ and
$S({\cal A})\tr{\cal A}-\det\cal A$ used to deduce the stability properties of
rolls. The matrix, the operator and the basis are expanded in a power series in
$\varepsilon$, whose coefficients depend on small parameters $\alpha$,
$\delta_x$ and $\delta_y$. Note that by virtue of (\ref{W02}),(\ref{W03})
the action of the mapping $(\delta_x,\delta_y)\to(-\delta_x,-\delta_y)$ amounts
to permutation of indices ${\bf W}_{2,0}\leftrightarrow{\bf W}_{3,0}$.
Consequently,
\begin{equation}
\renewcommand{\arraystretch}{1.6}
\begin{array}{l}
A_{12}(\delta_x,\delta_y)=A_{13}(-\delta_x,-\delta_y),\
A_{21}(\delta_x,\delta_y)=A_{31}(-\delta_x,-\delta_y),\\
A_{23}(\delta_x,\delta_y)=A_{32}(-\delta_x,-\delta_y),\
A_{22}(\delta_x,\delta_y)=A_{33}(-\delta_x,-\delta_y).
\end{array}
\label{sym}\end{equation}

Vector fields ${\bf W}_{j,0}$ (\ref{W01})-(\ref{W03}), representing terms
of order zero in $\varepsilon$ in the series (\ref{seriW}),
are eigenfunctions of $L_0$:
\begin{equation}
L_0{\bf W}_{j,0}=\lambda_{j,0}{\bf W}_{j,0},
\end{equation}
hence $A_{jj,0}=\lambda_{j,0}$ and $A_{ij,0}=0$ for $i\ne j$.
The following relations were established in OP2008:
$$
\renewcommand{\arraystretch}{1.6}
\begin{array}{l}
\lambda_{1,0}=-P(\delta_x^2+\delta_y^2),\\
\lambda_{j,0}=-C_4((k_\pm^2-k^2)^2+4\alpha k(k_\pm^2-k^2))
+\rO((k_\pm^2-k^2)^3,\alpha^2(k_\pm^2-k^2)),\ j=2,3,
\end{array}
$$
$$C_4=3P(4gk^2)^{-1},\quad g=\displaystyle\frac{1}{4}((2\pi^2-k^2)k^{-2}+3P).$$
Here and below in this Appendix, plus is assumed in place of $\pm$
for $j=2$, and minus for $j=3$.

The second and third terms of the series (\ref{seriL}) are
\begin{equation}
\label{L12}
\renewcommand{\arraystretch}{1.6}
\begin{array}{l}
L_1({\bf v},\theta)=(\Uvel_1\times(\nabla\times{\bf v})+
{\bf v}\times(\nabla\times\Uvel_1),
-(\Uvel_1\cdot\nabla)\theta-({\bf v}\cdot\nabla)\Utem_1),\\
L_2({\bf v},\theta)=(\Uvel_2\times(\nabla\times{\bf v})+
{\bf v}\times(\nabla\times\Uvel_2)+P\theta{\bf e}_z,
-(\Uvel_2\cdot\nabla)\theta-({\bf v}\cdot\nabla)\Utem_2).
\end{array}
\end{equation}

The $\varepsilon$ order entries of the matrix are calculated from the relation
\begin{equation}
L_0{\bf W}_{j,1}+L_1{\bf W}_{j,0}=\lambda_{j,0}{\bf W}_{j,1}+\sum_{i=1}^3 A_{ij,1}{\bf W}_{i,0}.
\label{w1}\end{equation}
Since the operator $L_0$ is self-adjoint with respect to the scalar product
\begin{equation}
\label{scpro}
({\bf w}_1,{\bf w}_2)={\bf w}_1^{\rm flow}\cdot{\bf w}_2^{\rm flow}
+PR_c{\bf w}_1^{\rm temp}\cdot{\bf w}_2^{\rm temp},
\end{equation}
the scalar product of (\ref{w1}) with ${\bf W}_{i,0}$ yields
$$A_{ij,1}=({\bf W}_{i,0},{\bf W}_{i,0})^{-1}(L_1{\bf W}_{j,0},{\bf W}_{i,0}),$$
which gives the $\rO(\varepsilon)$ terms of the matrix
$$
\renewcommand{\arraystretch}{1.8}
\begin{array}{l}
A_{j1,1}=\pm{1\over 2}kb\delta_y+bC_2\delta_x\delta_y+
\rO(\delta^3,\alpha\delta^2),\\
A_{1j,1}=(\delta_x^2+\delta_y^2)^{-1}
(-b\delta_x\delta_y\pi^2(2k^2)^{-1}\pm
b\pi^2(4k^3)^{-1}\delta_y(3\delta_x^2-\delta_y^2)+\rO(\delta^4,\alpha\delta^3)),
\end{array}
$$
where $j=2,3$ and $C_2=(Pk^2+\pi^2)(P+1)^{-1}(\pi^2+k^2)^{-1}$.
The remaining entries $A_{ij,1}$ vanish. We use the notation
$\rO(\delta^N)=\rO(\sum_{n=0}^N\delta_x^n\delta_y^{N-n})$.

Approximations to ${\bf W}_{j,1}$ for $j=2,3$ are also found from (\ref{w1}).
The O($\varepsilon^2$) entries are calculated using the equation
\begin{equation}
\label{w2}
L_0{\bf W}_{j,2}+L_1{\bf W}_{j,1}+L_2{\bf W}_{j,0}=
\lambda_{j,0}{\bf W}_{j,2}+\sum_{i=1}^3 A_{ij,1}{\bf W}_{i,1}+
\sum_{i=1}^3 A_{ij,2}{\bf W}_{i,0};
\end{equation}
the non-vanishing terms are
$$A_{jj,2}=-C_3\pm H_1(k\pi^2)^{-1}\delta_x+\rO(\delta^2,\alpha\delta),$$
$$A_{ji,2}=-C_3\pm H_2(k\pi^2)^{-1}\delta_x+\rO(\delta^2,\alpha\delta),$$
where $i,j=2,3$, $C_3=Pk^2(P+1)^{-1}a^{-2}$ and
$H_1+H_2=4P(27(P+1))^{-1}$ (in what follows only this sum is important).

Finally, entries of the matrix $\cal A$ are
\begin{equation}
\label{AA}
\renewcommand{\arraystretch}{1.6}
\begin{array}{lll}
A_{11}&=&-P(\delta_x^2+\delta_y^2)+\rO(\varepsilon^2\delta^2),\\
A_{21}&=&\displaystyle\frac{1}{2}kb\varepsilon\delta_y+\varepsilon bC_2\delta_x\delta_y+
\rO(\varepsilon\delta^3,\varepsilon\alpha\delta^2,\varepsilon^3),\\
A_{31}&=&-\displaystyle\frac{1}{2}kb\varepsilon\delta_y+\varepsilon bC_2\delta_x\delta_y+
\rO(\varepsilon\delta^3,\varepsilon\alpha\delta^2,\varepsilon^3),\\

A_{12}&=&
\varepsilon \biggl(-\displaystyle\frac{b\pi^2}{2k^2}\delta_x\delta_y
+\displaystyle\frac{b\pi^2}{4k^3}\delta_y(3\delta_x^2-\delta_y^2)\biggr)
(\delta_x^2+\delta_y^2)^{-1}+
\rO(\varepsilon\delta^2,\varepsilon\alpha\delta,\varepsilon^3),\\
A_{22}&=&-\varepsilon^2C_3+\varepsilon^2\displaystyle\frac{1}{k\pi^2}H_1\delta_x-C_4
\biggl((k_+^2-k^2)^2+4\alpha k(k_+^2-k^2)\biggr)\\
&&+\rO((k_+^2-k^2)^3,\alpha^2(k_+^2-k^2),\varepsilon^2\delta^2,\varepsilon^4),\\
A_{32}&=&-\varepsilon^2C_3+\varepsilon^2\displaystyle\frac{1}{k\pi^2}H_2\delta_x+
\rO(\varepsilon^2\delta^2,\varepsilon^4),\\

A_{13}&=&
\varepsilon \biggl(-\displaystyle\frac{b\pi^2}{2k^2}\delta_x\delta_y
-\displaystyle\frac{b\pi^2}{4k^3}\delta_y(3\delta_x^2-\delta_y^2)\biggr)
(\delta_x^2+\delta_y^2)^{-1}+
\rO(\varepsilon\delta^2,\varepsilon\alpha\delta,\varepsilon^3),\\
A_{23}&=&-\varepsilon^2C_3-\varepsilon^2\displaystyle\frac{1}{k\pi^2}H_2\delta_x+
\rO(\varepsilon^2\delta^2,\varepsilon^4),\\
A_{33}&=&-\varepsilon^2C_3-\varepsilon^2\displaystyle\frac{1}{k\pi^2}H_1\delta_x-C_4
\biggl((k_-^2-k^2)^2+4\alpha k(k_-^2-k^2)\biggr)\\
&&+\rO((k_-^2-k^2)^3,\alpha^2(k_-^2-k^2),\varepsilon^2\delta^2,\varepsilon^4).
\end{array}
\end{equation}

In the new basis $\widetilde{\bf W}_1={\bf W}_1$,
$\widetilde{\bf W}_2={\bf W}_2+{\bf W}_3$,
$\widetilde{\bf W}_3={\bf W}_2-{\bf W}_3$
the matrix of the operator $L$ is
\begin{equation}
\label{atil}
\renewcommand{\arraystretch}{1.6}
\begin{array}{lll}
\widetilde A_{11}&=&-P(\delta_x^2+\delta_y^2)+\rO(\varepsilon^2\delta^2),\\
\widetilde A_{21}&=&\varepsilon bC_2\delta_x\delta_y+
\rO(\varepsilon\delta^4,\varepsilon\alpha\delta^2,\varepsilon^3),\\
\widetilde A_{31}&=&\displaystyle\frac{1}{2}kb\varepsilon\delta_y+
\rO(\varepsilon\delta^3,\varepsilon^3),\\

\widetilde A_{12}&=&-\varepsilon
b\delta_x\delta_y\displaystyle\frac{\pi^2}{k^2}(\delta_x^2+\delta_y^2)^{-1}
+\rO(\varepsilon\delta^2,\varepsilon^3),\\
\widetilde A_{22}&=&-2\varepsilon^2C_3-C_4
(4k^2\delta_x^2+\delta_y^4+4\alpha k\delta_y^2)
+\rO(\alpha(\delta_x^2+\delta_y^4),\delta_x^4,\delta_y^6,\varepsilon^2\delta^2,
\varepsilon^4),\\
\widetilde A_{32}&=&-4C_4k\delta_x(\delta_y^2+2\alpha k)+
\varepsilon^2\displaystyle\frac{1}{k\pi^2}(H_1-H_2)\delta_x+
\rO(\alpha^2\delta_x,\delta_x^3,\delta_x\delta_y^4,\varepsilon^2\delta^3,\varepsilon^4\delta),\\

\widetilde A_{13}&=&\varepsilon
\displaystyle\frac{b\pi^2}{2k^3}\delta_y(3\delta_x^2-\delta_y^2)(\delta_x^2+\delta_y^2)^{-1}+
\rO(\varepsilon\delta^3,\varepsilon\alpha\delta,\varepsilon^3),\\
\widetilde A_{23}&=&-4C_4k\delta_x(\delta_y^2+2\alpha k)+
\varepsilon^2\displaystyle\frac{1}{k\pi^2}(H_1+H_2)\delta_x+
\rO(\alpha^2\delta_x,\delta_x^3,\delta_x\delta_y^4,\varepsilon^2\delta^3,\varepsilon^4\delta),\\
\widetilde A_{33}&=&-C_4(4k^2\delta_x^2+\delta_y^4+4\alpha k\delta_y^2)+
\rO(\alpha(\delta_x^2+\delta_y^4),\delta_x^4,\delta_y^6,\varepsilon^2\delta^2,\varepsilon^4).
\end{array}
\end{equation}
When calculating (\ref{atil}) with the use of (\ref{AA}), relations (\ref{sym})
were employed to estimate the omitted terms.

From (\ref{atil}) we obtain
\begin{equation}
\label{det0a}
\renewcommand{\arraystretch}{2.0}
\begin{array}{l}
\det{\cal A}=
d_{01}\delta_x^6+d_{02}\delta_x^4\delta_y^2+d_{03}\delta_x^2\delta_y^6+
d_{04}\delta_y^{10}\\
+\varepsilon^2(
\delta_x^2+\delta_y^2)^{-1}(d_{11}\delta_x^6+d_{12}\delta_x^4\delta_y^2+
d_{13}\delta_x^2\delta_y^4+d_{14}\delta_y^8)\\
+\varepsilon^4
(\delta_x^2+\delta_y^2)^{-1}(d_{21}\delta_x^2\delta_y^2+d_{22}\delta_y^4)
+\alpha(d_{31}\delta_x^2\delta_y^4+d_{32}\delta_y^8)\\
+\alpha^2(d_{41}\delta_x^4+d_{42}\delta_x^2\delta_y^2+d_{43}\delta_y^6)+
\alpha\varepsilon^2
(d_{51}\delta_x^2\delta_y^2(\delta_x^2+\delta_y^2)^{-1}+d_{52}\delta_y^4)\\
+\rO(\delta^2(\delta_x^2+\delta_y^4)^3,
\varepsilon^2\delta_x^2\delta^4,\varepsilon^2\delta_y^8,
\varepsilon^4\delta^4,\varepsilon^6\delta^2,\alpha\delta_y^4(\delta_x^2+\delta_y^4)^2,
\alpha^2\delta^2(\delta_x^2+\delta_y^4)^2,
\alpha\varepsilon^2\delta_y^2(\delta_x^2+\delta_y^4)),
\end{array}
\end{equation}
where
\begin{equation}
\label{ddd}
\renewcommand{\arraystretch}{2.2}
\begin{array}{l}
d_{01}=-\displaystyle\frac{16P^3}{(P+1)^2},\ d_{02}=d_{01},\
d_{03}=\displaystyle\frac{16P^3}{\pi^2(P+1)^2},\
d_{04}=-\displaystyle\frac{4P^3}{\pi^4(P+1)^2},\\
d_{11}=-\displaystyle\frac{16P^3}{9\pi^2(P+1)^2},\
d_{12}=\displaystyle\frac{32P(-3P^2+5P+1)}{27\pi^2(P+1)^2},\\
d_{13}=\displaystyle\frac{16P(-P^2+2P+2)}{9\pi^2(P+1)^2},\
d_{14}=\displaystyle\frac{-8P(P^2+2P+2)}{9\pi^4(P+1)^2},\\
d_{21}=\displaystyle\frac{224P}{243\pi^4(P+1)},\
d_{22}=-\displaystyle\frac{32P}{81\pi^4(P+1)},\
d_{31}=\displaystyle\frac{64P^3k}{\pi^2(P+1)^2},\\
d_{32}=-\displaystyle\frac{32P^3k}{\pi^4(P+1)^2},\
d_{41}=\displaystyle\frac{16P^3}{(P+1)^2},\ d_{42}=d_{41},\
d_{43}=-\displaystyle\frac{32P^3}{\pi^2(P+1)^2},\\
d_{51}=\displaystyle\frac{128Pk}{9\pi^2(P+1)},\
d_{52}=-\displaystyle\frac{32P(P^2+2P+2)k}{9\pi^4(P+1)^2};
\end{array}
\end{equation}

\begin{equation}
\label{min0a}
\renewcommand{\arraystretch}{1.6}
\begin{array}{l}
\tr{\cal A}S({\cal A})-\det{\cal A}=
e_{01}\delta_x^6+e_{02}\delta_x^4\delta_y^2+e_{03}\delta_x^2\delta_y^4+
e_{04}\delta_y^{8}\\
+\varepsilon^2
(\delta_x^2+\delta_y^2)^{-1}(e_{11}\delta_x^6+e_{12}2\delta_x^4\delta_y^2+
e_{13}\delta_x^2\delta_y^4+e_{14}\delta_y^6)+\\
\varepsilon^4
(\delta_x^2+\delta_y^2)^{-1}(e_{21}\delta_x^4+e_{22}\delta_x^2\delta_y^2+e_{23}\delta_y^4)+
\alpha^2 e_3\delta_x^4+
\alpha\varepsilon^2 e_4\delta_x^2\delta_y^2(\delta_x^2+\delta_y^2)^{-1}\\
+\rO(\delta^4(\delta_x^2+\delta_y^4)^2,
\varepsilon^2\delta^6,
\varepsilon^4\delta^4,\varepsilon^6\delta^2,\alpha\delta^4(\delta_x^2+\delta_y^4),
\alpha\varepsilon^2\delta_y^2(\delta_x^2+\delta_y^4)),
\end{array}
\end{equation}
where
\begin{equation}
\label{eee}
\renewcommand{\arraystretch}{2.2}
\begin{array}{l}
e_{01}=-\displaystyle\frac{8P^3(P+5)^2}{(P+1)^3},\
e_{02}=-\displaystyle\frac{16P^3(P+5)}{(P+1)^2},\
e_{03}=-\displaystyle\frac{8P^3}{(P+1)},\
e_{04}=-\displaystyle\frac{4P^3}{\pi^2(P+1)},\\
e_{11}=-\displaystyle\frac{4P^3(P^2+18P+65)}{9\pi^2(P+1)^3},\
e_{12}=\displaystyle\frac{4P(-9P^4-54P^3-99P^2+20P+2)}{27\pi^2(P+1)^3},\\
e_{13}=-\displaystyle\frac{4P(3P^3+3P^2+22P+26)}{27\pi^2(P+1)^2},\
e_{14}=-\displaystyle\frac{4P^2(P^2+2P+4)}{9\pi^2(P+1)},\\
e_{21}=-\displaystyle\frac{16P^3(P+5)}{81\pi^4(P+1)^3},\
e_{22}=-\displaystyle\frac{32P(P+3)(3P^2+2P+2)}{273\pi^4(P+1)^3},\\
e_{23}=-\displaystyle\frac{16P^3}{81\pi^4(P+1)^2},\
e_3=-\displaystyle\frac{512P^3}{(P+1)^3},\
e_4=-\displaystyle\frac{128Pk}{9\pi^2(P+1)}.
\end{array}
\end{equation}

\section{A necessary and sufficient condition for existence of
eigenvalues of a $3\times3$ matrix, which have positive real parts}

In this Appendix we show that instead of direct calculation of eigenvalues,
in order to study stability of rolls it suffices to check, whether
any of the inequalities (\ref{zeroe}) or (\ref{ime}) is satisfied for
some $\delta_x$ and $\delta_y$.

We start by exposition of three lemmas
about eigenvalues of a $3\times3$ matrix.

\medskip\noindent
{\it Lemma 1.} Let $\cal A$ be a $3\times3$ matrix with real entries. Denote
its eigenvalues by $\lambda_i$, $i=1,2,3$, and the sum of the second order
minors by $S({\cal A})$:
$$S({\cal A})=A_{11}A_{22}-A_{12}A_{21}+A_{11}A_{33}-A_{13}A_{31}+
A_{22}A_{33}-A_{23}A_{32}.$$
Consider the following statements:

$$
\renewcommand{\arraystretch}{1.6}
\begin{array}{l}
\hbox{\bf S1: } \displaystyle{\max_{1\le i\le 3}}(\hbox{Re}\lambda_i)>0\\
\hbox{\bf S2: } \det{\cal A}>0\\
\hbox{\bf S3: } S({\cal A})\tr{\cal A}-\det{\cal A}>0
\end{array}
$$

\medskip\noindent
Then
$$(i)\ {\bf S2}\Rightarrow{\bf S1}\ \hbox{ and }\ (ii)\ {\bf S3}\Rightarrow{\bf S1}.$$

\medskip\noindent
{\it Proof.} $(i)$ follows from the identity
$\det{\cal A}=\lambda_1\lambda_2\lambda_3$ (consider separately two cases:
all eigenvalues are real, or two of them are complex conjugate).

Similarly, $(ii)$ follows from the identities
$$S({\cal A})=\lambda_1\lambda_2+\lambda_1\lambda_3+\lambda_2\lambda_3,\quad
\tr{\cal A}=\lambda_1+\lambda_2+\lambda_3$$
(again, consider separately the two cases).

\medskip
If in the condition of the Lemma the signs "$>$" are replaced by "$\ge$",
the statements remain true. The modified lemma is referred to as Lemma 1'.

\medskip\noindent
{\it Lemma 2.} Let $\cal A({\bf x})$ be a $3\times3$ matrix with real entries
continuously depending on ${\bf x}\in\Omega\subset{\bf R}^n$, where $\Omega$ is
a connected domain in ${\bf R}^n$. Denote by $\lambda_i({\bf x})$,
$i=1,2,3$ the eigenvalues of $\cal A({\bf x})$. Suppose
$$
\renewcommand{\arraystretch}{1.6}
\begin{array}{l}
\exists\ {\bf x}_0\in\Omega\quad\displaystyle{\max_{1\le i\le 3}}\hbox{Re}\lambda_i({\bf x}_0)<0,\\
\det{\cal A}\ne0\quad \forall{\bf x}\in\Omega,\\
S({\cal A})\tr{\cal A}-\det{\cal A}\ne0,\quad \forall{\bf x}\in\Omega.
\end{array}
$$

\medskip\noindent
Then
$$\max_{1\le i\le 3}\hbox{Re}\lambda_i({\bf x})<0\quad\forall{\bf x}\in\Omega.$$

\medskip\noindent
{\it Proof.} Suppose there exists ${\bf x}_1\in\Omega$ such that an
eigenvalue of $\cal A({\bf x})$, say, $\lambda_1$, has a positive real part.
A curve in $\Omega$ connects ${\bf x}_0$ and ${\bf x}_1$. The eigenvalue
$\lambda_1$ is a continuous function on this curve (because roots of the cubic
equation $\det({\cal A}-\lambda I)=0$
are continuous functions of its coefficients.) Since $\hbox{Re}\lambda_1$ has
different signs at ${\bf x}_0$ and ${\bf x}_1$, there exists a point
$\hat{\bf x}$ on the curve such that $\hbox{Re}\lambda_1(\hat{\bf x})=0$.
If $\lambda_1(\hat{\bf x})=0$, then $\det{\cal A}(\hat{\bf x})=0$, and if
$\lambda_1(\hat{\bf x})$ is imaginary, then
$S({\cal A}(\hat{\bf x}))\tr{\cal A}(\hat{\bf x})-\det{\cal A}(\hat{\bf x})=0$,
since in this case
$$S({\cal A})\tr{\cal A}-\det{\cal A}=\hbox{Re}\lambda_1(2\lambda_3^2+
4\hbox{Re}\lambda_1\lambda_3+2\lambda_1\bar\lambda_1)=
2\hbox{Re}\lambda_1((\lambda_3+\hbox{Re}\lambda_1)^2+
(\hbox{Im}\lambda_1)^2).$$
Thus a contradiction with the statement of the lemma is obtained and the lemma
is proved.

\medskip\noindent
{\it Lemma 3.} Let $\cal A({\bf x})$ and $\Omega$ be the same as in the
statement of Lemma 2. Assume
$$\quad\exists\ {\bf x}_0\in\Omega\quad
\displaystyle{\max_{1\le i\le 3}}\hbox{Re}\lambda_i({\bf x}_0)<0,$$
\begin{equation}
\label{detn0}
\quad\det{\cal A}\ne0\quad \forall{\bf x}\in\Omega,
\end{equation}
$$\quad\exists\ {\bf x}_1\in\Omega\quad S({\cal A}({\bf x}_1))\tr{\cal A}({\bf x}_1)-
\det{\cal A}({\bf x}_1)>0.$$

\medskip\noindent
Then
$$\exists\ {\bf x}_2\in\Omega\quad \hbox{Re}\lambda_1({\bf x}_2)>0,
\ \hbox{Im}\lambda_1({\bf x}_2)\ne0.$$

\medskip\noindent
{\it Proof.} A curve in $\Omega$ connects ${\bf x}_0$ and ${\bf x}_1$.
Let $\xi$ be a parameter along this curve, $\xi=0$ at
${\bf x}_0$ and $\xi=1$ at ${\bf x}_1$. There exist
$\xi_0$, $0<\xi_0<1$, and $\xi_1$, $\xi_0<\xi_1<1$, such that
\begin{equation}
\label{zero}
S({\cal A}(\xi_0))\tr{\cal A}(\xi_0)-\det{\cal A}(\xi_0)=0,
\end{equation}
$$S({\cal A}(\xi))\tr{\cal A}(\xi)-\det{\cal A}(\xi)>0\quad\forall\ \xi\in(\xi_0,\xi_1],$$
$$S({\cal A}(\xi))\tr{\cal A}(\xi)-\det{\cal A}(\xi)\le0\quad\forall\ \xi\in[0,\xi_0].$$

Suppose all eigenvalues of ${\cal A}(\xi_0)$ are real. By Lemma 1' all of them
are non-positive in $[0,\xi_0]$. Due to (\ref{detn0}) they do not vanish
at $\xi_0$, hence they are strictly negative at $\xi_0$.
But then (\ref{zero}) can not be satisfied. Hence the assumption
that all eigenvalues are real is wrong.

Let $\lambda_1$ and $\lambda_2=\overline\lambda_1$ be a pair of complex
eigenvalues and $\lambda_3$ be real. By continuity, there exists $\xi_2$,
$\xi_2>\xi_0$, such that $\hbox{Im}\lambda_1(\xi)\ne0$ for all
$\xi\in[\xi_0,\xi_2]$. The expression
$$S({\cal A})\tr{\cal A}-\det{\cal A}=
2\hbox{Re}\lambda_1((\lambda_3+\hbox{Re}\lambda_1)^2+
(\hbox{Im}\lambda_1)^2)$$
is positive only if $\hbox{Re}\lambda_1$ is positive.
Consequently, $\hbox{Re}\lambda_1(\xi_3)>0$ and
$\hbox{Im}\lambda_1(\xi_3)\ne0$. The lemma is proved.

\medskip
The Lemmas are applied to investigate stability of rolls.

Let $\cal A$ be the matrix calculated in Appendix A. Assume
$(\delta_x,\delta_y)$ is the parameter $\bf x$ employed in Lemma 2,
$\varepsilon$, $\alpha$ and $P$ being fixed.
If (\ref{zeroe}) or (\ref{ime}) is satisfied for some $(\delta_x,\delta_y)$,
Lemma 1 implies existence of an eigenvalue with positive real part.

Suppose (\ref{zeroe}) and (\ref{ime}) are not satisfied for any $\delta_x$
and $\delta_y$. For sufficiently
large $\delta_x$ ($\delta_x^2\gg\varepsilon^2$ and $\delta_x^2\gg\alpha^2$)
the matrix has three real negative eigenvalues. Let $\Omega$ be
${\bf R}^2$ with the origin excluded. Conditions of Lemma 2
are satisfied, hence for any $(\delta_x,\delta_y)$ all the eigenvalues
have negative real parts.

Note that (\ref{zeroe}) implies that the matrix $\cal A$ has a real
eigenvalue with a positive real part, while (\ref{ime}) does not
guarantee that there exist a pair of complex eigenvalues with a
positive real part. However, assume in addition that $\det{\cal A}<0$ for all
$(\delta_x,\delta_y)$ (or for all $(\delta_x,\delta_y)$ in a
connected region $\Omega$, where conditions of Lemma 3 are satisfied).
Then by Lemma 3 there exists a point $(\delta_x,\delta_y)$
where $\cal A$ has a complex eigenvalue with a positive real part.

\section{A bound for $\alpha<0$, $P>P_1$, $\alpha^2\sim\varepsilon^4$}

We prove here that under the conditions, stated in the title of the Appendix,
\begin{equation}
\label{appB}
D_0+\varepsilon^2D_1+\varepsilon^4d_{22}\delta_y^4(\delta_x^2+\delta_y^2)^{-1}
+\alpha D_3+\alpha^2D_4+\alpha\varepsilon^2d_{52}\delta_y^4
\end{equation}
is negative for all $\delta_x$ and $\delta_y$.

To begin with, note that the terms
involving $\alpha$ are asymptotically small and therefore are neglected.

Assume $\delta_y^2\gg\delta_x^2$. In the leading order (\ref{appB}) is
\begin{equation}
\label{detB1}
\delta_y^2(d_{02}\delta_x^4+\delta_x^2(\varepsilon^2d_{13}+
d_{03}\delta_y^4)+\varepsilon^4d_{22}+\varepsilon^2d_{14}\delta_y^4
+d_{04}\delta_y^8).
\end{equation}
This is a quadratic polynomial in $\delta_x^2$, which admits a maximum at
\begin{equation}
\label{detB2}
\delta_x^2=-{{d_{03}\delta_y^4+\varepsilon^2d_{13}}\over2d_{02}}.
\end{equation}
The maximum is
$$-{\delta_y^2\over 4d_{02}}\biggl(\varepsilon^2(2d_{03}d_{13}-4d_{02}d_{14})\delta_y^4+
\varepsilon^4(d_{13}^2-4d_{02}d_{22})\biggr),$$
where both expressions in the brackets are negative for $P>P_1$.

Now assume $\delta_x^2\sim\delta_y^2$ or $\delta_x^2\gg\delta_y^2$.
In the leading order (\ref{appB}) is equal to
\begin{equation}
\label{detB3}
(d_{01}\delta_x^6+d_{02}\delta_x^4\delta_y^2)+
\varepsilon^2
(d_{11}\delta_x^6+d_{12}\delta_x^4\delta_y^2+d_{13}\delta_x^2\delta_y^4)(\delta_x^2+\delta_y^2)^{-1}+
\varepsilon^4d_{22}\delta_y^4(\delta_x^2+\delta_y^2)^{-1}.
\end{equation}
This quadratic polynomial in $\varepsilon^2$ can take positive values only if
\begin{equation}
\label{detB4}
d_{11}\delta_x^6+d_{12}\delta_x^4\delta_y^2+d_{13}\delta_x^2\delta_y^4>0.
\end{equation}
If this is satisfied, the maximum (in $\varepsilon^2$) of (\ref{detB3}) is
\begin{equation}
\label{detB5}
-(d_{11}\delta_x^6+d_{12}\delta_x^4\delta_y^2+
d_{13}\delta_x^2\delta_y^4)^2(4d_{22}\delta_y^4)^{-1}+d_{01}\delta_x^8
+(d_{01}+d_{02})\delta_x^6\delta_y^2+d_{02}\delta_x^4\delta_y^4.
\end{equation}
In view of the inequalities (\ref{detB4}), $d_{11}<0$ and $d_{13}>d_{12}$ for $P>P_1$,
$$d_{13}\delta_x^4\delta_y^2+d_{13}\delta_x^2\delta_y^4>d_{11}\delta_x^6+d_{12}\delta_x^4\delta_y^2+
d_{13}\delta_x^2\delta_y^4.$$
Note that $d_{01}=d_{02}$ and $d_{13}^2<4d_{22}d_{02}$ for $P>P_1$; hence
(\ref{detB5}) is always negative for $P>P_1$.

\section{The large $P$ limit}

In this Appendix we calculate an instability boundary, which is
important for $\alpha<0$ and large $P$.
In the limit of large $P$ the coefficient $f_2$ in (\ref{f2}) vanishes
and instability occurring for $\alpha^2\sim\varepsilon^2$ may compete
with the instability defined by (\ref{f2}).

Suppose $\alpha^2\sim\varepsilon^2$. Then $\det{\cal A}>0$ for $\alpha<0$,
if either $\delta_x^2\gg\delta_y^2$ or $\delta_y^2\gg\delta_x^2$
(due to the presence of the term $\alpha\varepsilon^2D_5$).

Suppose $\delta_y^2\gg\delta_x^2$. Represent $\det\cal A$ (\ref{det0a}) (it
is simpler to calculate this directly from (\ref{AA})~) as
$$\det{\cal A}=$$
\begin{equation}
-\delta_y^2(2\varepsilon^2C_3+C_4\xi)(\varepsilon^2b^2{\pi^2\over4k^2}+PC_4\xi)
\label{lp1}\end{equation}
\begin{equation}
-2PC_4k^4\delta_x^4\delta_y^2+\alpha\varepsilon^2b^2\pi^2k\delta_x^2
\label{lp2}\end{equation}
\begin{equation}
+\delta_x^2\delta_y^2(-PC_4k^2(\varepsilon^2C_3+C_4\xi)+
4P\alpha^2k^2+{\varepsilon^2b^2\pi^2\over2}),
\label{lp3}\end{equation}
where
$$\xi=\alpha k\delta_y^2+\delta_y^4/4.$$
For $\delta_x=0$ the determinant is given by (\ref{lp1}). Considering
$$\varepsilon^2b^2{\pi^2\over4k^2}+PC_4\xi$$
as a quadratic polynomial in $\delta_y^2$, we find that (\ref{lp1})
is positive (we are interested in large $P$'s, and for them
the instability boundary is defined by the second term in (\ref{lp1}) ) for
\begin{equation}
\varepsilon^2<f_3\alpha^2,\quad f_3={9\pi^2P^2\over2(P+1)}.
\label{ip4}\end{equation}
Note that
\begin{equation}
\lim_{P\to\infty}f_3=\infty.
\label{lim4}\end{equation}

For large $P$ and $\varepsilon$ not satisfying (\ref{ip4}), the contributions
to $\det\cal A$ from (\ref{lp2}) and (\ref{lp3}) are negative (the proof is
omitted).
Hence if $\det{\cal A}<0$ for $\delta_x=0$ and all
$\delta_y$, it remains negative for all $\delta_x$ and $\delta_y$.
Thus (\ref{ip4}) is indeed a boundary for stability of rolls.

For $\delta_x^2\gg\delta_y^2$ the instability boundary is that
of the Eckhaus instability
$$\varepsilon^2<36\alpha^2\pi^2$$
(or in a more familiar form $R-R^s<3(R_c(k)-R^s)$\,),
which is below the boundary defined by (\ref{ip4}).

\bigskip
\makeatletter
\renewcommand*{\@biblabel}[1]{}
\makeatother

\pagebreak
\centerline{
\psfig{file=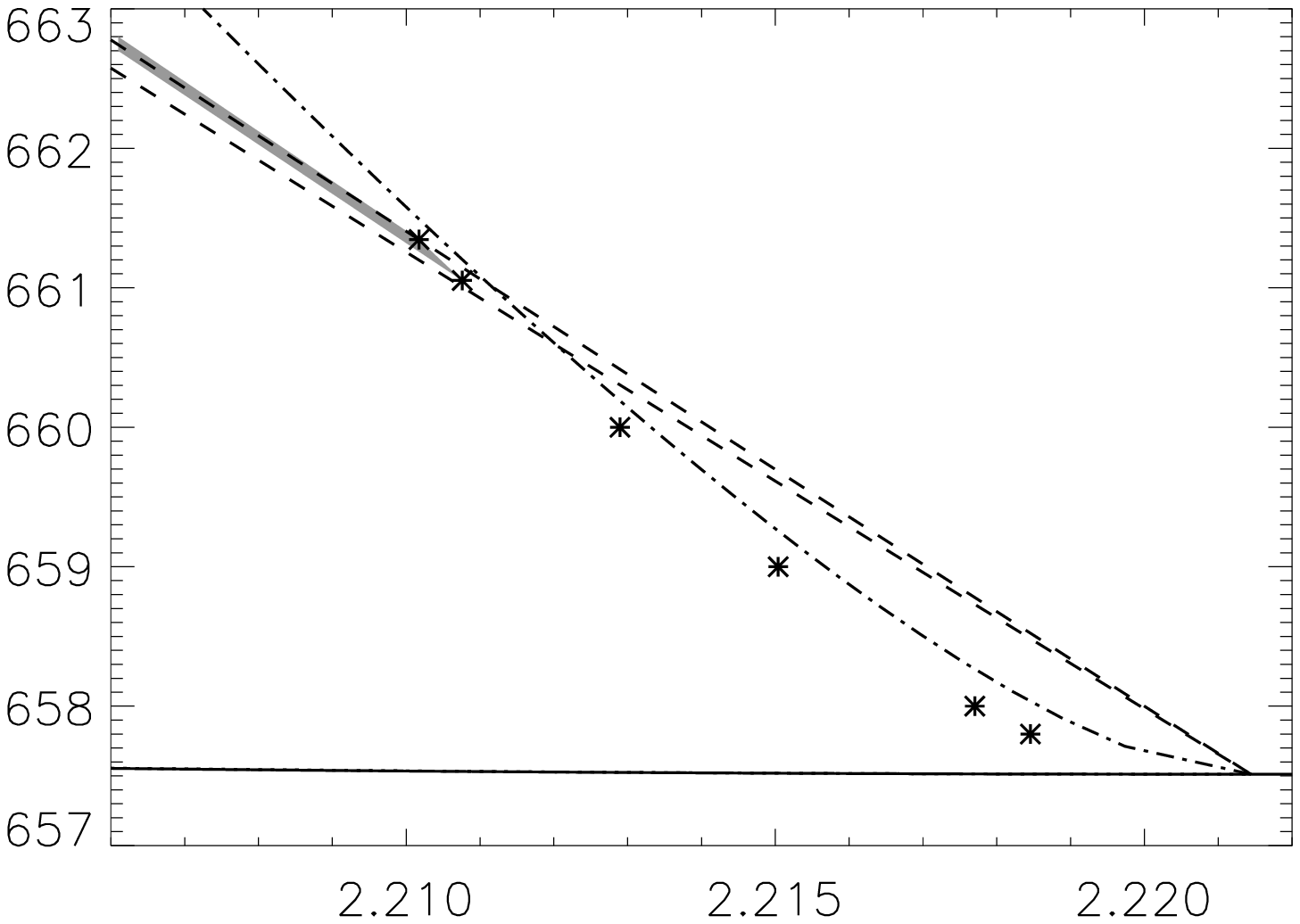,width=80mm}~
\psfig{file=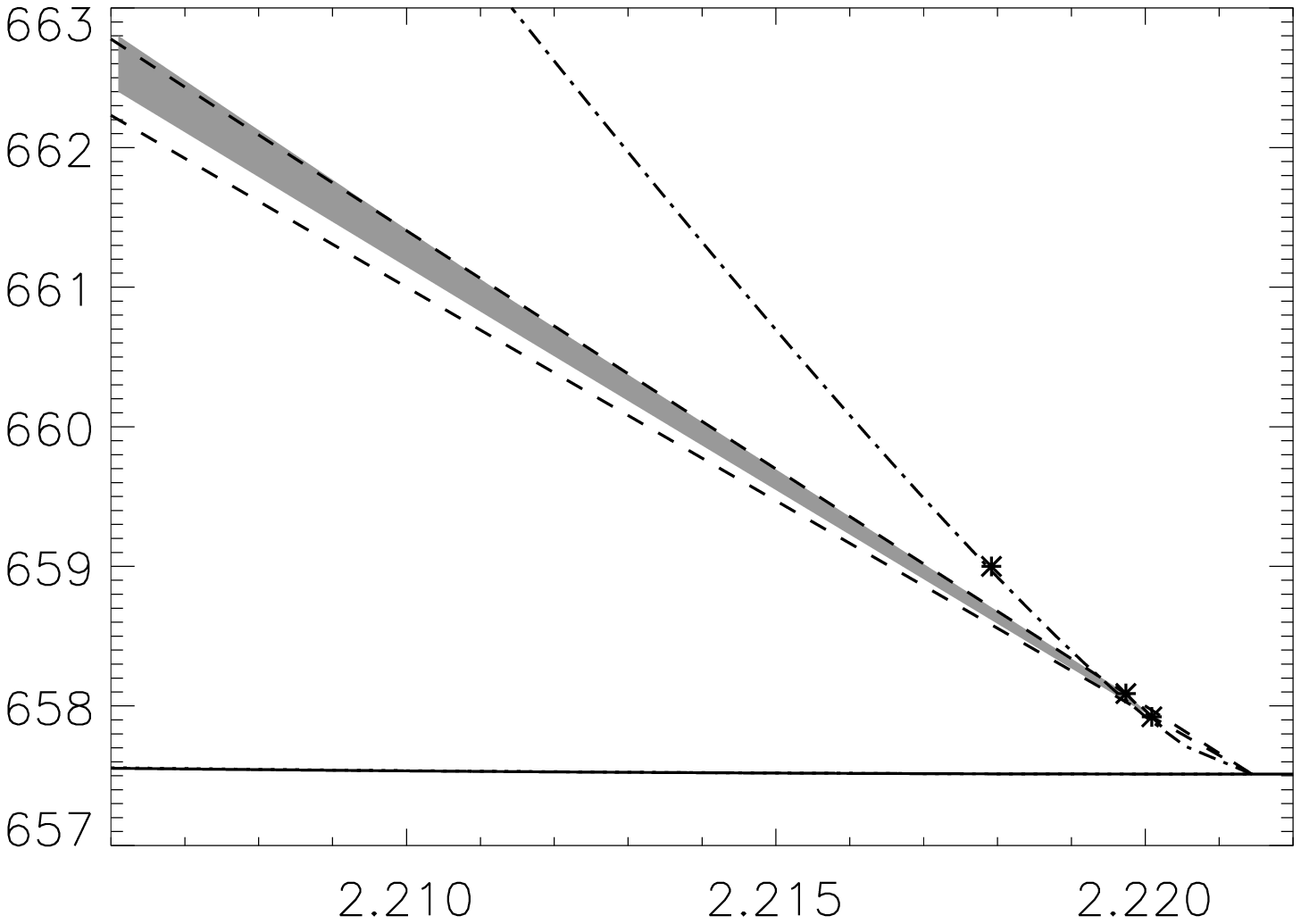,width=80mm}
}

\vspace{-.2cm}
\centerline{
(a)\hspace{65mm}(b)
}

~

\centerline{
\psfig{file=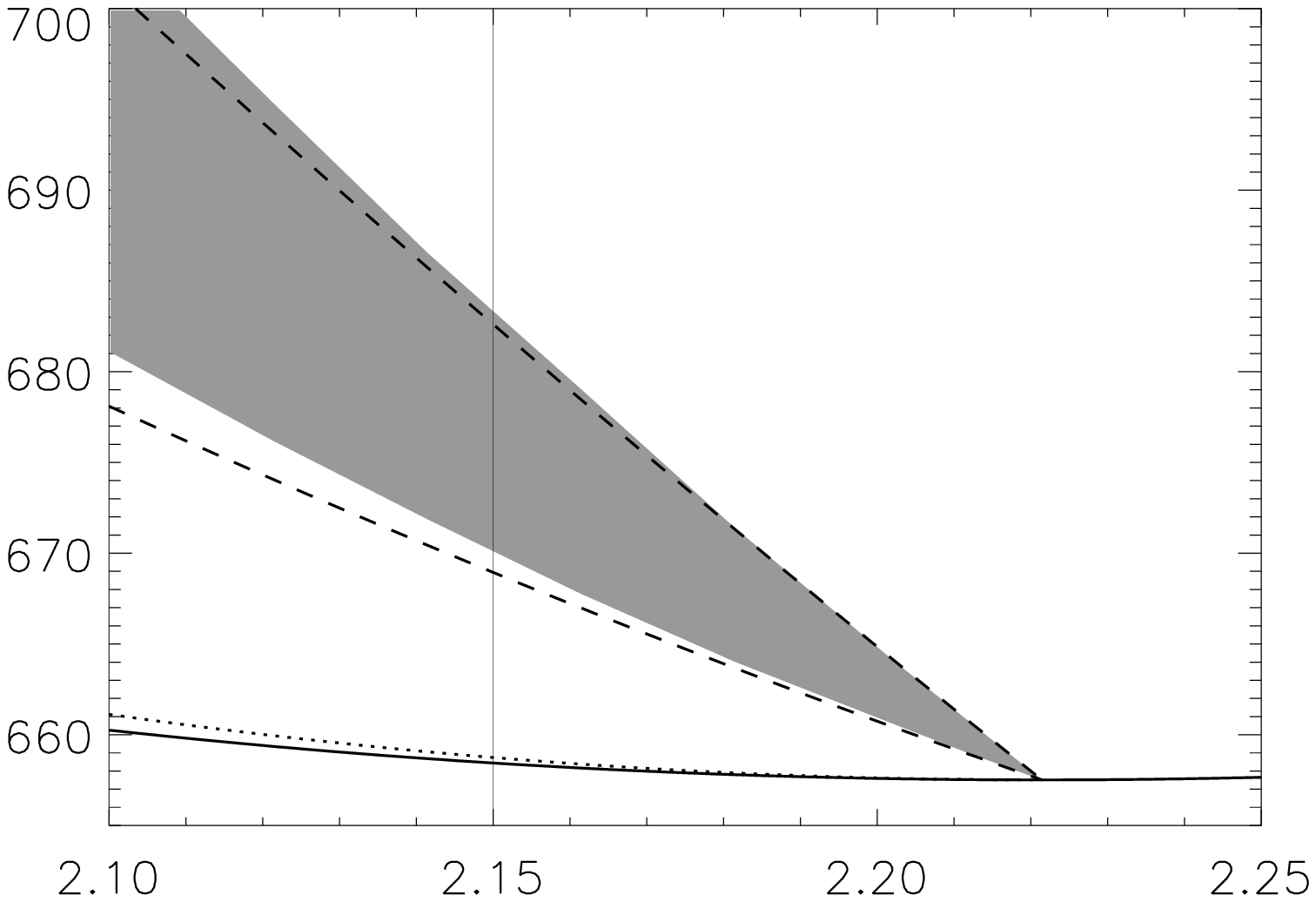,width=80mm}~
\psfig{file=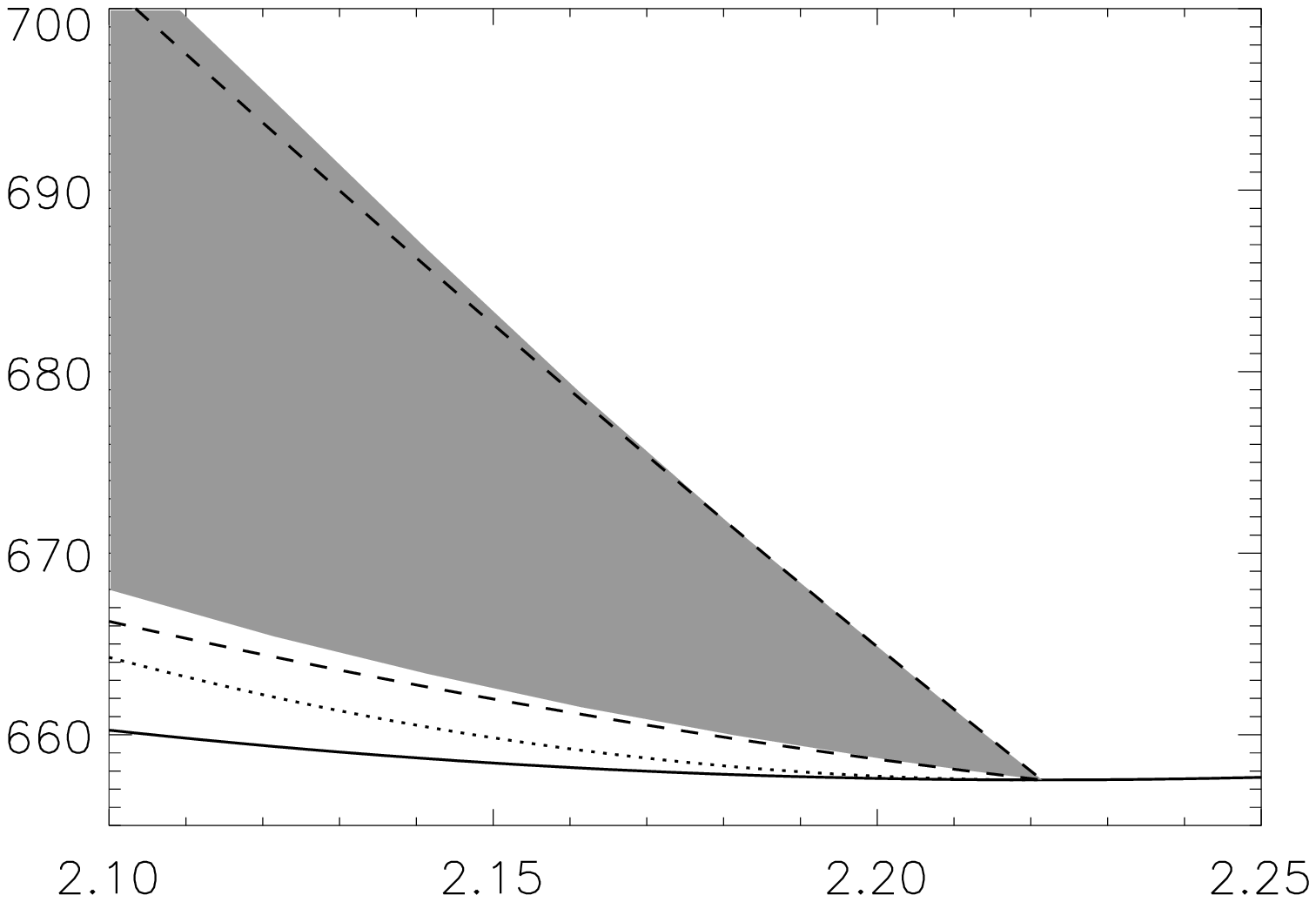,width=80mm}
}

\vspace{-.2cm}
\centerline{
(c)\hspace{65mm}(d)
}
~

\centerline{
\psfig{file=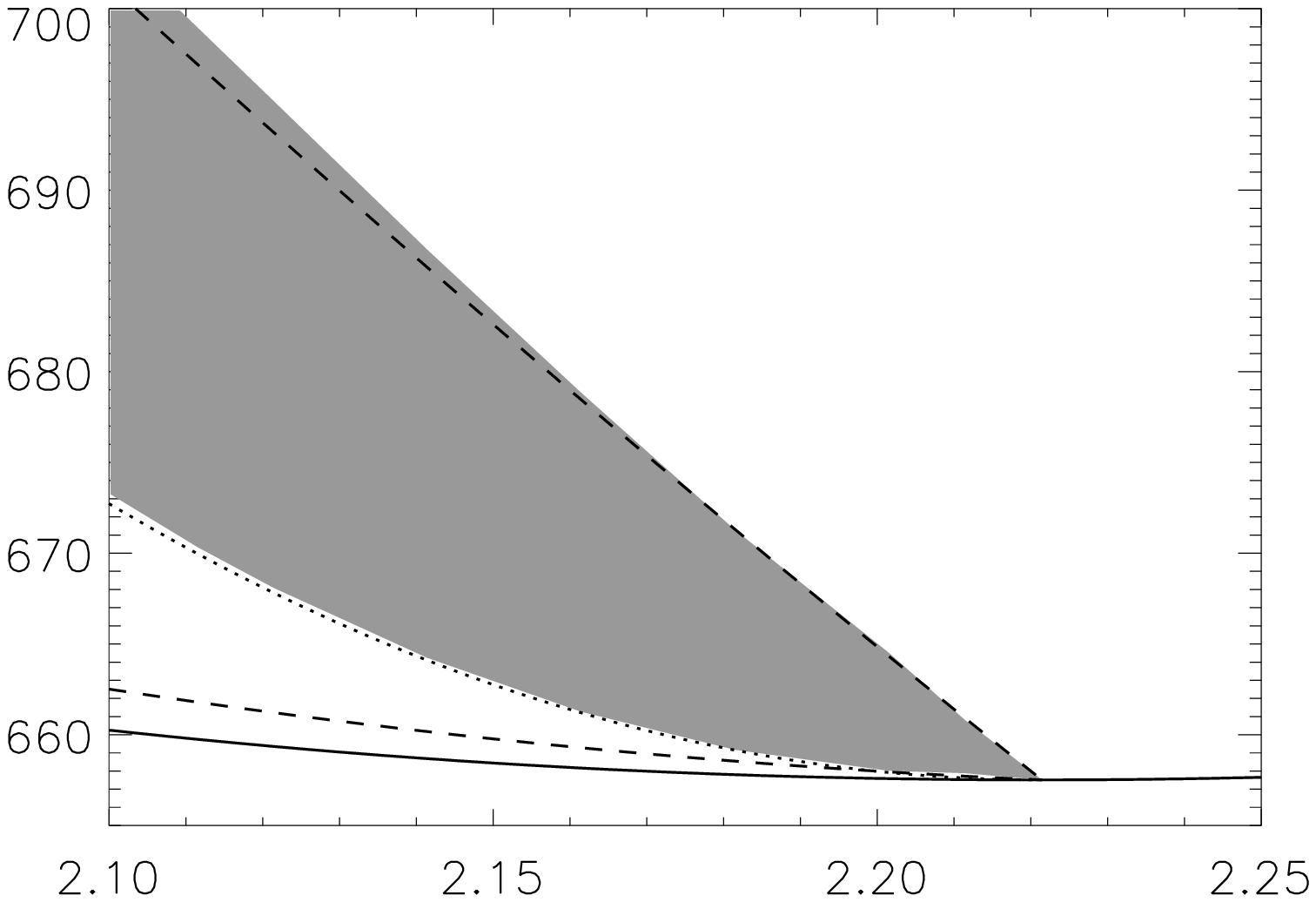,width=80mm}~
\psfig{file=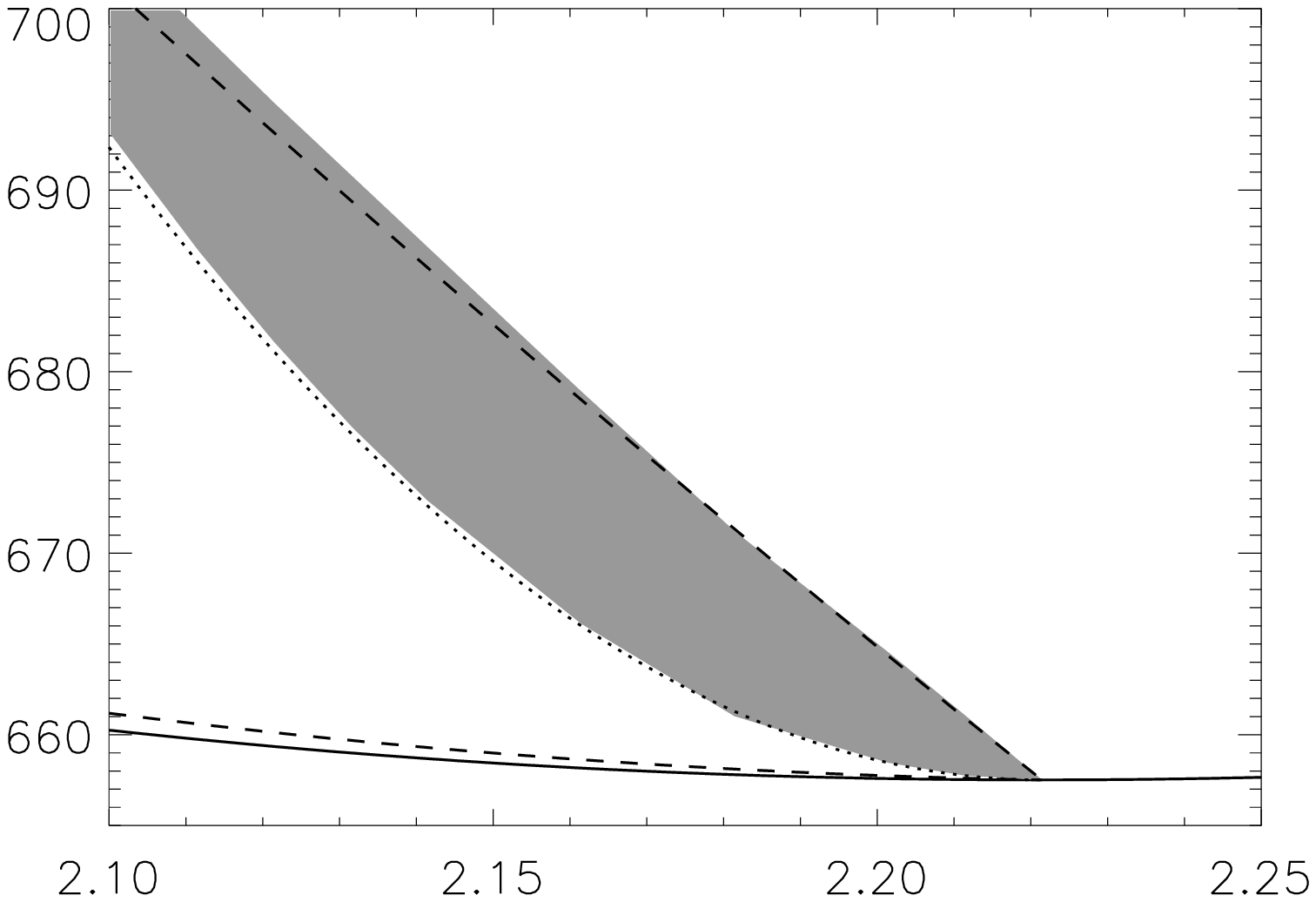,width=80mm}
}

\vspace{-.2cm}
\centerline{
(e)\hspace{65mm}(f)
}

\vspace{0.5cm}
\noindent
Figure 1. The area of stable rolls (shaded) on the $(k,R)$ plane found
numerically (see section 8) and the instability boundaries found analytically
for $P=0.6$ (a) $P=0.7$ (b), $P=2$ (c) and $P=7$ (d), $P=20$ (e)
and $P=50$ (f). Solid line denotes the onset of convection, dashed
lines instability boundaries SV and OSV defined by (\ref{f1}) and
(\ref{f2}), dotted line the ZZ boundary (\ref{ipn}) and
dashed-dotted line the SV2 boundary (\ref{f4n}), (\ref{f4nn}).
Stars mark the points where the SV2 boundary is obtained by interpolation.
Horizontal axis: $k$, vertical axis: $R$.

\vspace*{-218mm}\hspace*{5mm}$\bf OSV$

\vspace*{-14mm}\hspace*{14mm}$\bf SV2$

\vspace*{11mm}\hspace*{37mm}$\bf SV$

\vspace*{-11mm}\hspace*{87mm}$\bf OSV$

\vspace*{-16mm}\hspace*{111mm}$\bf SV2$

\vspace*{-5mm}\hspace*{93mm}$\bf SV$

\vspace*{60mm}\hspace*{13mm}$\bf SV$

\vspace*{28mm}\hspace*{12mm}$\bf ZZ$

\vspace*{-13mm}\hspace*{4mm}$\bf OSV$

\vspace*{-30mm}\hspace*{96mm}$\bf SV$

\vspace*{31mm}\hspace*{100mm}$\bf ZZ$

\vspace*{-9mm}\hspace*{86mm}$\bf OSV$

\vspace*{28mm}\hspace*{13mm}$\bf SV$

\vspace*{24mm}\hspace*{4mm}$\bf ZZ$

\vspace*{3mm}\hspace*{14mm}$\bf OSV$

\vspace*{-42mm}\hspace*{96mm}$\bf SV$

\vspace*{28mm}\hspace*{100mm}$\bf OSV$

\vspace*{-22mm}\hspace*{87mm}$\bf ZZ$

\pagebreak

\centerline{
\psfig{file=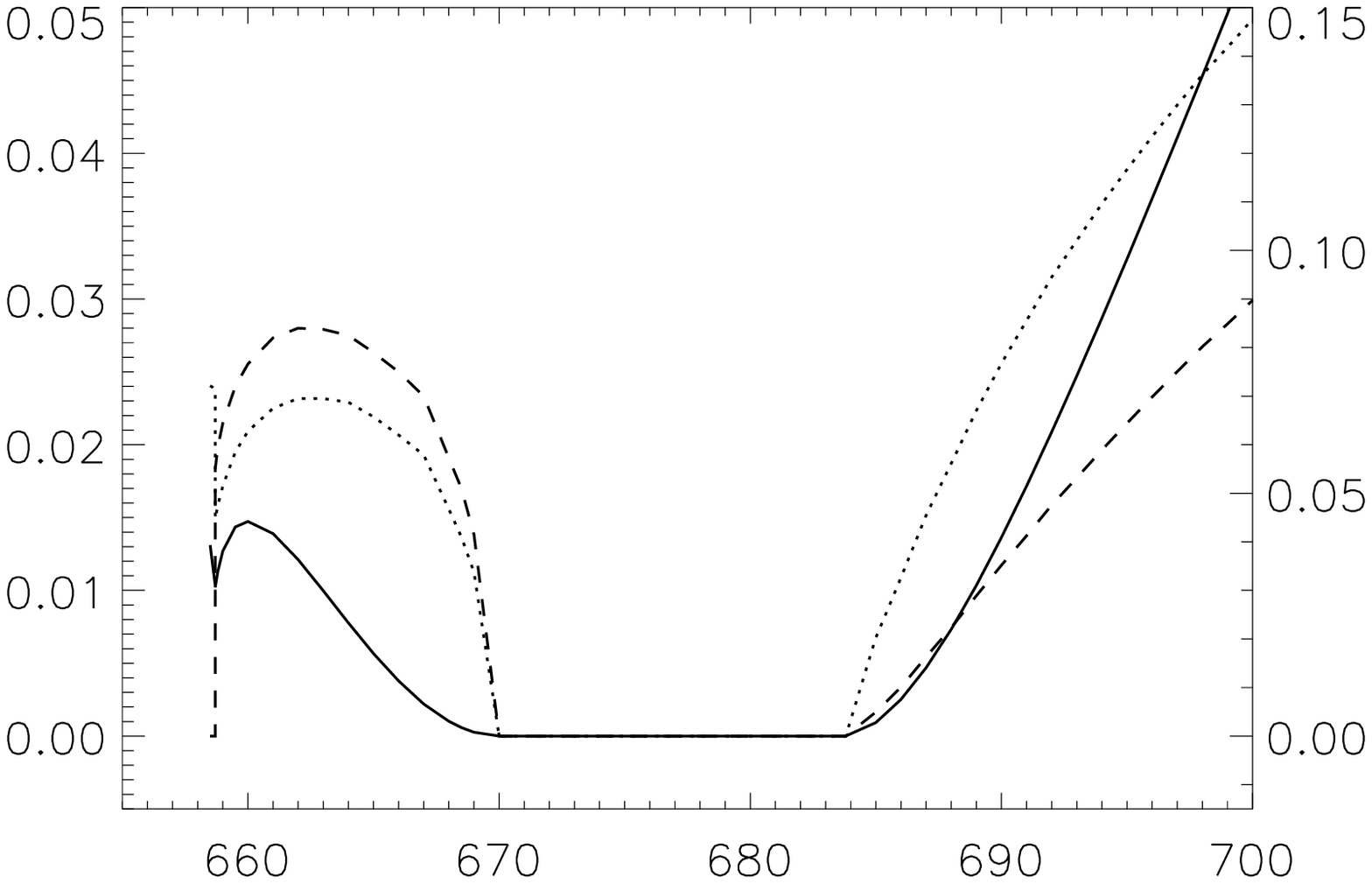,width=150mm}
}

\vspace{1cm}
\noindent
Figure 2. The dominant growth rate (solid line, left vertical axis) and
the values of $\delta_x$ and $\delta_y$ (dotted and dashed lines, respectively,
right axis), where the maximum is achieved, versus the Rayleigh number
(horizontal axis) for $P=2$ and $k=2.15$ (the respective crossection
is shown by a thin vertical line on fig.~1c).

\end{document}